\documentclass[final,10pt,3p,twocolumn]{elsarticle}

\journal{Journal of Magnetism and Magnetic Materials}

\usepackage[utf8]{inputenc}
\usepackage[T1]{fontenc}
\usepackage{amsmath}
\usepackage{amssymb}
\usepackage{graphicx}
\usepackage[skip=0pt]{subcaption}
\usepackage[format=plain,labelfont=bf]{caption}
\usepackage[version=3]{mhchem}
\usepackage[per-mode=symbol,range-phrase=--,range-units=single]{siunitx}
\usepackage{booktabs}
\usepackage{todonotes}
\usepackage{hyperref}
\usepackage{cleveref}

\hypersetup{
  colorlinks=true,
}

\makeatletter
\providecommand{\doi}[1]{%
  \begingroup
    \let\bibinfo\@secondoftwo
    \urlstyle{rm}%
    \href{http://dx.doi.org/#1}{%
      doi:\discretionary{}{}{}%
      \nolinkurl{#1}%
    }%
  \endgroup
}
\makeatother

\crefname{equation}{Eq.}{Eqs.}
\crefname{section}{Sec.}{Secs.}
\crefname{table}{Tab.}{Tabs.}
\crefname{figure}{Fig.}{Figs.}
\crefname{subfigure}{Fig.}{Figs.}

\presetkeys{todonotes}{inline}{}

\biboptions{sort&compress}

\graphicspath{{figures/}}

\usepackage{threeparttable}

\newcommand*{\partic}{\mathrm{p}}
\newcommand*{\liq}{\mathrm{\ell}}
\newcommand*{\vap}{\mathrm{v}}
\newcommand*{\ferrofl}{\mathrm{f}}
\newcommand*{\basefl}{{\mathrm{bf}}}

\newcommand{\textcite}[1]{\citet{#1}}

\makeatletter
\newcommand*{\dd}{\@ifnextchar^{\DIfF}{\DIfF^{}}}
\def\DIfF^#1{\mathop{\mathrm{\mathstrut d}}\nolimits^{#1}\gobblesp@ce}
\def\gobblesp@ce{\futurelet\diffarg\opsp@ce}
\def\opsp@ce{%
  \let\DiffSpace\!%
  \ifx\diffarg(%
    \let\DiffSpace\relax
  \else
    \ifx\diffarg[%
      \let\DiffSpace\relax
    \else
      \ifx\diffarg\{%
        \let\DiffSpace\relax
      \fi\fi\fi\DiffSpace}
\makeatother

\newcommand*{\ddx}[1]{\ensuremath{\frac{\dd #1}{\dd x}}}
\newcommand*{\pdt}[1]{\ensuremath{\frac{\partial #1}{\partial t}}}
\newcommand*{\pdx}[1]{\ensuremath{\frac{\partial #1}{\partial x}}}
\newcommand*{\pd}[3][]{\ensuremath{\frac{\partial^{#1} {#2}}{\partial{#3}^{#1}}}}
\newcommand*{\vct}[1]{\ensuremath{\boldsymbol{#1}}}
\renewcommand*{\vec}[1]{\ensuremath{\boldsymbol{#1}}}

\renewcommand*{\Re}{\text{Re}}

\begin{document}

\begin{frontmatter}



\title{A multi-phase ferrofluid flow model with equation of state for
  thermomagnetic pumping and heat transfer}


\author[ER]{Eskil Aursand\corref{cor}}
\ead{eskil.aursand@sintef.no}
\author[ER]{Magnus Aa. Gjennestad}
\author[ER]{Karl Yngve Lerv{\aa}g}
\author[ER]{Halvor Lund}


\address[ER]{SINTEF Energy Research, P.O. Box 4761 Sluppen, NO-7465 Trondheim, Norway}
\cortext[cor]{Corresponding author. (Tel.: +4793005349)}

\begin{abstract}
A one-dimensional multi-phase flow model for thermomagnetically
pumped ferrofluid with heat transfer is proposed. 
The thermodynamic model is a combination of a simplified particle
model and thermodynamic equations of state for the base fluid. 
The magnetization model is based on statistical mechanics, taking into
account non-uniform particle size distributions. 
An implementation of the proposed model is validated against experiments
from the literature, and found to give good predictions for the 
thermomagnetic pumping performance. However, the results reveal a very large
sensitivity to uncertainties in heat transfer coefficient predictions.\\

\small{
\textcopyright~2015. 
This manuscript version is made available under the 
\href{http://creativecommons.org/licenses/by-nc-nd/4.0/}{CC-BY-NC-ND 4.0}
license.

\doi{10.1016/j.jmmm.2015.11.042}
}

\end{abstract}

\begin{keyword}



Heat transfer \sep 
Ferrofluid \sep 
Thermomagnetic pump\sep
Fluid mechanics \sep
Equation of state

\end{keyword}

\end{frontmatter}


\section{Introduction}
\label{sec:introduction}
Heat exchange is of key significance in a number of applications, such
as process design and integration, 
waste heat recovery and collection, household heating and cooling,
and cooling of engines, electronics and power electronics. Heat
exchange concepts often use a fluid as the means of heat transport, and the
rate of heat transfer to and from the fluid is a limitation.

In 1995, \textcite{Choi95} proposed adding nanoparticles to a
fluid to enhance its heat transfer properties.
The particles are normally smaller than
\SI{100}{\nano\meter}, which ensures that they are suspended in the
fluid by Brownian agitation. Surfactants are typically also added to
improve the stability of the particle suspension. The term nanofluid is used
for fluids that consist of a base fluid, such as water, oil or glycol,
with suspended nanoparticles and possibly added surfactants.

While the theoretical potential of nanofluids has been known for some
time, steadily improving nanofabrication techniques are opening more
and more possibilities for practical use. Nanofluids have been shown
to have quite novel properties, such as increased thermal conductivity 
and Nusselt number~\cite{Kakacc09, Xuan00_2, Maiga05}.
They have been heavily researched for the last 10--20 years,
and a wide range of potential applications have been proposed 
(see e.g.~\textcite{Taylor13}).

If a nanofluid is synthesized with magnetic particles, it is known as
a ferrofluid \cite{Odenbach03}. Such magnetic nanofluids open up the
possibility of pumping the fluid using an inhomogeneous magnetic field.
A pump utilizing magnetic fields would require fewer moving parts, perhaps
none whatsoever, which may lead to increased reliability.

Due to the symmetry of a static magnetic field, such a field cannot
alone produce a net force on the fluid in steady-state, 
so another effect is needed to break the symmetry. The effect used here
is the fact that the magnetic susceptibility of a ferrofluid depends on
temperature. If the particles are engineered to have a particularly strong
response to temperature, the fluids are called 
temperature-sensitive magnetic fluids (TSMF).
A commonly used TSMF is composed of Mn--Zn
ferrite particles with different kinds of base fluids~\cite{Arulmurugan06}.

The concept of using magnetically pumped ferrofluids, often referred
to as thermomagnetic pumping, for heat exchange has been demonstrated
by a number of authors, see e.g.~\textcite{Lian09a}, \textcite{Xuan11}
and \textcite{Lee12}. \textcite{Iwamoto11} built an apparatus for
measuring the net driving force of a thermomagnetic pump for different
heat rates and pipe inclinations.

In this paper, we present a model for multi-phase ferrofluid flow
that includes the effects of applied heat and a magnetic field on the
fluid. We include a thermodynamic equation of state and vapor--liquid
equilibrium calculations in order to accurately predict the
thermodynamic properties of the base fluid. 
Equations of state enable the prediction of thermodynamic
properties in a consistent way~\cite{Michelsen07},
across a wide range of pressures and temperatures, and common implementations
include parameter databases for a large variety of possible mixtures. 
Taking advantage of the results of this large field of research adds
flexibility to the model.
To the best of our knowledge,
including such thermodynamic models is novel work when it comes to
simulation of thermomagnetic pumping. 

An implementation of the model is then validated by
comparing simulation results with the experimental results by
\textcite{Iwamoto11}.

In \cref{sec:model}, the equations for 
one-dimensional ferrofluid flow are
presented, along with the source terms for magnetic, 
frictional, and gravitational forces, as well as heat transfer.
The thermodynamic model and magnetization
equations are also described. \Cref{sec:numerics} briefly
explains the numerical methods used to solve the model equations. 
The
validation of the model against experimental results is described in
\cref{sec:validation}.
In \cref{sec:discussion} we
discuss the results, and finally we draw conclusions and outline further
work in \cref{sec:conclusions_and_further_work}.


\section{Model}
\label{sec:model}

In this section, we present our multi-phase ferrofluid flow model. The
model consists of a set of one-dimensional conservation equations with
source terms that model the effects of heat transfer, friction,
gravity, and magnetic forces. The equations are closed by a
thermodynamic model that relates the primary flow variables to an
equilibrium state. In addition, we present a model for the dependency
of the ferrofluid magnetization on the magnetic field and
temperature. The development of the flow model builds on ferrofluid
dynamics models by \textcite{Rosensweig85,Muller01,Tynjala05},
though with significant simplifications expected to be appropriate for
these applications.

\subsection{Fluid description}
\label{sub:fluid_description}

We wish to describe the multi-phase flow (liquid/vapor/nanoparticles)
of a ferrofluid in a pipe under the influence of external forces and 
heat sources/sinks. In this work, the system is described by a one-dimensional 
homogeneous equilibrium multi-phase flow model. 
In such a one-dimensional description, the actual sharp boundaries
between phases are not resolved. Instead, the multi-phase state at a
given position along the pipe is described by the volume fraction of
each phase at that position.  The local volume fraction of phase $k$
is given by $\alpha_k$ (--).  The index $k$ may be used to describe
the three main phases, or different unions of them, as summarized in
\cref{tab:phase_indices}.  The sum of the main volume fractions is
unity, i.e.~$\alpha_\liq + \alpha_\vap + \alpha_\partic = 1$.

Similarly, each phase has its own local average density, 
$\rho_k$~(\si{\kilogram\per\meter\cubed}), which combines to the 
mixture density $\rho$ in the following way:
\begin{equation}
  \rho = \sum_k \alpha_k \rho_k \quad (k \in \{\ferrofl,\liq,\partic\}).
  \label{eq:total_rho_from_sum}
\end{equation}
The density is also defined for the combined phase indices $\ferrofl$ 
and $\basefl$, as the combined mass divided by the combined volume. 

The volume fractions and densities of each phase must be found through some 
thermodynamic model, given the main flow-variables: 
mass fluxes, pressure, and enthalpy. 
A central assumption in enabling this is the 
homogeneous equilibrium model (HEM), which means that 
chemical, thermal and mechanical equilibriums between phases are 
locally reached instantaneously. 
In other words, it is assumed that chemical potential, temperature and pressure
are equal in all phases at any given time and position 
(though they may vary in time and space). The model also
includes the assumption that the friction between the phases is large enough
to make the velocities equal.

The base fluid itself may consist of several chemical components, as is 
often the case with working fluids in heat transfer systems. 
However, in the homogeneous equilibrium model, the total composition of the 
base fluid will be constant. The compositions of the liquid and the vapor phase
may vary, and be different from each other and the total composition, but this
is not relevant for the flow model, which only requires the densities and 
phase fractions. However, the local compositions are relevant for 
the underlying thermodynamic model.

\begin{table}[tbp]
  \centering
  \caption{Description of phase index subscripts.}
  \label{tab:phase_indices}

  \begin{tabular}{ll}
    \toprule
    Phase index     & Description       \\
    \midrule
    $k$           & Generic phase index \\
    $\liq$        & Liquid phase \\
    $\vap$        & Vapor phase \\
    $\partic$     & Particle phase \\
    $\ferrofl$    & Ferroliquid phase ($\liq$ and $\partic$) \\
    $\basefl$     & Base fluid phase ($\liq$ and $\vap$) \\
    No subscript  & The combined $\liq$ + $\vap$ + $\partic$ system \\
    \bottomrule
  \end{tabular}
\end{table}

\subsection{Flow equations}
\label{sub:flow_equations}

\subsubsection{Transient}
\label{ssub:transient}

By considering the conservation of particle mass, base fluid mass, 
total linear momentum and total energy, while considering source terms 
deemed relevant, we may derive one-dimensional transient flow equations. 
These are essentially the fluid dynamic Euler equations with added 
source terms, which in this case become
\begin{gather}
  \pdt{} \left(\alpha_\partic \rho_\partic \right) 
  + \pdx{} \left(\alpha_\partic \rho_\partic v \right) = 0,
  \label{eq:flow_transient_pmass}
  \\
  \pdt{} \left(\alpha_\basefl \rho_\basefl \right) 
  + \pdx{} \left(\alpha_\basefl\rho_\basefl v \right) = 0,
  \label{eq:flow_transient_bfmass}
  \\
  \pdt{} \left(\rho v\right) + \pdx{}\left(\rho v^2 + p\right)   
  = f^\text{mag} + f^\text{fric} + f^\text{grav},
  \label{eq:flow_transient_mom}\\
  \begin{split}
    \pdt{} \left( \rho e + \frac{1}{2} \rho v^2 \right) 
    + \pdx{} \left( v \left(\rho e + \frac{1}{2} \rho v^2 + p\right)\right)&
    \\
    =  v f^\text{grav} + \dot{q},&
  \end{split}
  \label{eq:flow_transient_energy}
\end{gather}
where~\cref{eq:flow_transient_pmass} represents the conservation of particles,
\cref{eq:flow_transient_bfmass} represents the conservation of 
base fluid (liquid + vapor) mass,
\cref{eq:flow_transient_mom} represents the conservation of total momentum,
and~\cref{eq:flow_transient_energy} represents the conservation of 
total (internal + kinetic) energy.
Here 
$x$~(\si{\meter}) is the position along the pipe, 
$t$~(\si{\second}) is time,
$v$~(\si{\meter\per\second}) is the flow velocity,
$p$~(\si{\pascal}) is the pressure
and
$e$~(\si{\joule\per\kilogram}) is the combined specific internal energy. 

The terms on the right-hand side are 
\emph{force terms}~(\si{\newton\per\meter\cubed}) and
the \emph{heat transfer term}~(\si{\watt\per\meter\cubed}),
commonly called \emph{source terms}, which will be explained in 
\cref{sub:source_terms}. 
The reason why the frictional force term is not present in the energy 
equation is that friction does not affect the total energy, it only
converts kinetic energy to internal energy. In principle there is also
an energy exchange with the fluid system when the magnetization 
changes (magnetocaloric effect), but
this will locally be negligible compared to the other source terms, and 
in steady-state, the energy exchange will sum to zero across the 
thermomagnetic pump as a whole. It is assumed that this term will have a
negligible effect on the results, 
and thus this term is excluded for model simplicity.
Additional implicit assumptions in these equations include no mass transfer
to or from the particle phase and no diffusion of particles.

\subsubsection{Steady state}
\label{ssub:steady_state}
Equations for steady state (stationary) flow may be derived
from~\crefrange{eq:flow_transient_pmass}{eq:flow_transient_energy}
by setting the time-differentiated terms equal to zero. 
We may then rewrite the momentum and 
energy equations into simpler equations for
pressure and specific total enthalpy $h$~(\si{\joule\per\kilogram}),
\begin{equation}
  h \equiv e + \frac{p}{\rho},
  \label{eq:enthalpy_def}
\end{equation}
by using the above definition and the conservation of 
total mass ($\rho v$ is constant), to yield 
\begin{gather}
  \ddx{} \left( \alpha_\partic \rho_\partic v \right) = 0,
  \label{eq:flow_steady_pmass}
  \\
  \ddx{} \left( \alpha_\basefl \rho_\basefl v \right) = 0,
  \label{eq:flow_steady_bfmass}
  \\
  \ddx{} \left( p \right) = - v \rho \ddx{}\left( v \right) 
  + f^{\text{mag}} + f^{\text{fric}} + f^{\text{grav}},
  \label{eq:flow_steady_p}
  \\
  \ddx{}\left( \rho v h \right) = - \frac{\rho v}{2}\ddx{}
  \left( v^2 \right)
  + v f^{\text{grav}}
  + \dot{q}.
  \label{eq:flow_steady_h}
\end{gather}

The above is a set of differential equations for four left hand side 
variables, which we will call the \emph{primary flow variables}.

The first two equations, \cref{eq:flow_steady_pmass} and
\cref{eq:flow_steady_bfmass}, simply state that the quantities 
$\alpha_\partic \rho_\partic v$ (particle mass flux)
and
$\alpha_\basefl \rho_\basefl v$ (base fluid mass flux) are constant
along the pipe. 
The last two equations, \cref{eq:flow_steady_p} and
\cref{eq:flow_steady_h}, are equations for 
$p$ (pressure) and $\rho v h$ (enthalpy flux), which would 
need to be integrated along the pipe to yield the varying values. 

It is important to note that the equations listed so far are not sufficient
to solve the system. A thermodynamic model, together with 
the assumption of instantaneous thermodynamic equilibrium, is needed to 
close the system.
The right-hand sides of \cref{eq:flow_steady_p,eq:flow_steady_h} 
depend on other quantities than the
ones on the left-hand side, such as temperature, densities and 
volume fractions. This is where the thermodynamic equilibrium 
calculations come in, which essentially find the
equilibrium state given a set of primary flow variables: 
\begin{equation}
  \begin{split}
  \text{Thermodynamics} :
    \left(
      \alpha_\partic \rho_\partic v,
      \alpha_\basefl \rho_\basefl v,
      p,
      \rho v h
    \right) 
  \\
  \mapsto
    \left(
      T,
      \rho_\liq,
      \rho_\vap,
      \rho_\partic,
      \alpha_\liq,
      \alpha_\vap,
      \alpha_\partic
    \right)
  \label{eq:thermodynamics}
\end{split}
\end{equation}
The process in~\cref{eq:thermodynamics} is covered in 
\cref{sub:thermodynamics}.
The primary flow variables, together with the output of the 
equilibrium calculation, can then be used to construct the right-hand
sides of \cref{eq:flow_steady_p,eq:flow_steady_h}.

\subsection{Source terms}
\label{sub:source_terms}

\subsubsection{Magnetic force}
\label{ssub:magnetic_force}

The magnetic force source term 
$f^\text{mag}$~(\si{\newton\per\meter\cubed})
is present in 
the pressure equation~\eqref{eq:flow_steady_p}, 
and represents 
the force acting on a fluid element with magnetization
$\vct{M}$~(\si{\ampere\per\meter}) caused by 
a magnetic field $\vct{H}$~(\si{\ampere\per\meter}).

We make the assumption that the magnetization $\vct{M}$ is always in
equilibrium with the $\vct{H}$-field. The most important magnetization
relaxation process for ferrofluid particles is Brownian relaxation,
which takes place on time scales of the order of \SI{e-4}{\second}
\cite{Muller01}, which is much smaller than the time scales of the
flow, and thus justifies the equilibrium assumption. We also neglect
magnetostrictive effects, which are important only if the particles
are compressible \cite{Rosensweig85}.

With these assumptions, the magnetic force on the magnetic particles
is the \emph{Kelvin magnetic force} on magnetized materials, which
states that the force acting on a small dipole (e.g.~small piece of
magnetized material) in an inhomogeneous magnetic field is given by
\cite{Rosensweig85}

\begin{equation}
  \vct{f}^{\text{Kelvin}} =
  \mu_0 \left( \vct{M} \cdot \nabla \right) \vct{H}_\text{ex},
  \label{eq:force_kelvin}
\end{equation}
where $\mu_0$~(\si{\newton\per\ampere\squared}) is the magnetic
permeability of vacuum. The Kelvin force is sometimes given in other
forms, such as $\vct{M} \cdot \nabla \vct{H}_\text{ex}$, but these are
equivalent given that $\nabla \times \vct{H}_\text{ex} = 0$.
Maxwell's equations state that this is true if there are no free
currents and no time-varying electric fields.

The field $\vct{H}_\text{ex}$ is the field external to the dipole, i.e.~the
field that would be present at that location if the dipole were not
there. This is sensible, since the field added by the dipole itself cannot 
contribute to the total force acting on the dipole, due to 
conservation of momentum.
In terms of the application in this work, $\vct{H}_\text{ex}$ must formally not
be confused with the field which would be present if none 
of the magnetizable fluid was present, $\vct{H}_0$. The field added by
other fluid elements may certainly affect the force applied to the 
fluid element in question. However, $\vct{H}_0$ may be a good 
approximation for $\vct{H}_\text{ex}$, and in
\cref{sub:solenoid_magnetic_field} we will argue that this 
is the case here.

For the case of a pipe of magnetizable material inside a solenoid 
electromagnet of a much larger diameter, 
the magnetic field has two special properties:
First, the axial component $H_x$ dominates over the other components.
Second, $H_x$ does not vary much radially within the pipe. 
When we also assume that the magnetization is isotropic 
($M$ and $H$ are collinear),
these approximations allow the reduction of~\cref{eq:force_kelvin} 
to a one-dimensional form:
\begin{equation}
  f^{\text{Kelvin}} = 
  \mu_0 M \pdx{H_{\text{ex}}}.
  \label{eq:force_kelvin_1D}
\end{equation}

The force term in the one-dimensional flow model is the total force
per volume applied to the infinitesimally thin cross section of the
ferroliquid plus vapor system at a given location. The expression
in~\cref{eq:force_kelvin_1D} is the force applied to a small volume
of magnetizable material, and would thus need to be integrated over
the cross section. Since only the ferroliquid contains magnetic
particles and has a magnetization $M_\ferrofl$, the force
density is
\begin{equation}
  f^\text{mag} = \frac{1}{A} \int_A f^{\text{Kelvin}} \dd{A} =
  \alpha_\ferrofl \mu_0 M_\ferrofl \pdx{H_\text{ex}},
  \label{eq:force_term_mag}
\end{equation}
where $A$ is the pipe cross-sectional area. Magnetization is a
volumetric property (dipole moment per volume), so the factor $M =
\alpha_\ferrofl M_\ferrofl$ can be seen as the average magnetization
of the total cross section.

An important point may be illustrated by considering the expected
pressure increase in a long straight pipe with a magnetizable fluid of
constant density with the general response $M=\chi H_\text{ex}$, when
passing a solenoid placed at $x=0$, given that it is only affected by
the magnetic force:
\begin{align}
  \Delta p &= \int_{-\infty}^{\infty} f^\text{mag} \dd{x} 
  \nonumber \\
            &= -\frac{\mu_0 }{2}
              \int_{-\infty}^{\infty}   \pdx{\chi}  H_\text{ex}^2 \dd{x}. 
\end{align}
From the above, one can see that if the magnetization response is 
constant in space ($\partial \chi / \partial x = 0$), no net pressure 
increase will be achieved. This is not surprising, as a static magnetic field
can not do work on its own. The symmetry must be broken by some 
inhomogeneity, which will supply the free energy for pumping. 
For the cases studied here, the general response $\chi$ is a function
of both temperature and the local amount of magnetizable particles, 
and these may both be used to break the symmetry.

\subsubsection{Frictional force}
\label{ssub:friction_force}
The frictional force source term $f^{\text{fric}}$~(\si{\newton\per\meter\cubed})
is present in the pressure equation~\cref{eq:flow_steady_p}, and represents
the momentum loss in the fluid cross section as a whole due to the no-slip
condition at the pipe side walls.

To approximate this term, we use the standard 
Darcy--Weisbach equation for frictional losses in pipes,
\begin{equation}
  f^{\text{fric}} = - \frac{f_D \rho v |v|}{2D},
  \label{eq:force_term_fric}
\end{equation}
where $D$~(\si{\meter}) is the pipe diameter, and 
$f_D$~(--) is the \emph{Darcy friction factor}, 
which may depend on flow characteristics. 
For perfectly laminar flow, $f_D = 64/\Re$, 
which reduces the frictional force to 
\begin{equation}
  f^{\text{fric}}_\text{laminar} =  - \frac{32 \eta v}{D^2},
  \label{eq:force_term_fric_laminar}
\end{equation}
where $\eta$~(\si{\pascal\second}) is the dynamic viscosity of the fluid.

%

In the case of liquid--vapor flow, the above is not strictly valid. 
However, in the case of equal phase velocities and well-mixed bubbly flow, 
one can reasonably make the approximation of treating the fluid as 
a \emph{pseudo single-phase} 
fluid, applying the single phase models with a mixture 
viscosity~\cite[Sec.~2.3.2]{Collier94}. 


\subsubsection{Gravitational force}
\label{ssub:gravitational_force}
The gravitational force source term 
$f^\text{grav}$~(\si{\newton\per\meter\cubed})
is present in both 
the pressure equation~\eqref{eq:flow_steady_p} and 
the enthalpy equation~\eqref{eq:flow_steady_h}, and represents 
the axial component of the gravitational force acting on a fluid element.
It is given by
\begin{equation}
  f^{\text{grav}} = - \rho g \sin (\theta),
  \label{eq:force_term_grav}
\end{equation}
where $g$~(\si{\meter\per\second\squared}) is the
gravitational acceleration constant, 
and $\theta$~(--) is the local inclination angle of the 
positive direction along the pipe compared to the horizontal.

\subsubsection{Wall heat transfer}
\label{ssub:heat_transfer}
The wall heat transfer source 
term $\dot{q}$~(\si{\watt\per\meter\cubed})
is present in the 
enthalpy equation~\eqref{eq:flow_steady_h}, and represents the
heat transfer rate between the pipe side walls and the fluid.

The heat transfer rate per fluid volume
for a fluid element flowing through a pipe 
in contact with a side-wall of temperature difference 
$\Delta T_\mathrm{wall}$~(\si{\kelvin}) is
\begin{equation}
  \dot{q} = \frac{4}{D} U \Delta T_\mathrm{wall},
  \label{eq:source_term_heat}
\end{equation}
where $U ~(\si{\watt\per\meter\squared\per\kelvin})$ is the \emph{heat
  transfer coefficient} (HTC).

%
%

\subsection{Thermodynamics}
\label{sub:thermodynamics}
In this section, we describe the procedure that lies behind the 
mapping~\eqref{eq:thermodynamics}.
Our basic assumptions are that there is instantaneous relaxation to thermal,
mechanical, chemical and magnetic equilibrium in the ferrofluid. This means that
we have the same temperature, pressure and chemical potential of base-fluid
components in all phases, and that we can use equilibrium models to
calculate fluid properties and particle magnetization.

\subsubsection{Base fluid}
\label{ssub:base_fluid}

To model the thermodynamic properties of the base fluid, we use an 
equation of state~\cite{Michelsen07}. The overall model should be compatible
with any one, but in this work we choose to employ the
Lee--Kesler equation of state \cite{Lee75}, which offers better density
predictions than the more common cubic equations, without being too 
computationally expensive. 
In this work, we assume that the base fluid may
split into at most two phases, vapor and liquid. As our base fluid consists
purely of hydrocarbons, this is a reasonable assumption.

To find the phase equilibrium state of the base fluid, we specify its
temperature $T$, pressure $p$
and chemical composition $\vec{z}$ (--). The equilibrium state is then found by
solving
a system of non-linear equations, expressing material balance of each chemical
component between the phases, that the chemical potential of each component is
the same in both phases, and that the temperature and pressure
are equal to the specified values.
We assume that the presence of nanoparticles in the liquid phase does not
significantly disturb the chemical potentials of the base fluid components.

After having found the phase equilibrium state, the equation of state allows us
to calculate e.g.\ $\rho_\liq$, $\rho_\vap$, $\rho_\basefl$ and $h_\basefl$. The
entire operation of finding the phase equilibrium and calculating the required
quantities can be thought of as the
mapping
\begin{equation}
  \text{Base fluid equil.} : \left( T, p, \vec{z} \right)
  \mapsto \left( \rho_\basefl, h_\basefl, \rho_\liq, \rho_\vap \right).
  \label{eq:basefluidtpflash}
\end{equation}
For a thorough discussion of how this is done, the reader is referred to the
work of \textcite{Michelsen07}.

\subsubsection{Particles}
\label{ssub:particles}

The thermodynamic properties of the particles are defined by a of
constant density $\rho_\partic$ and a constant specific heat capacity
at constant pressure $c^p_{\partic}$\
(\si{\joule\per\kelvin\per\kilo\gram}). Since the particles are solid,
their thermal expansion is negligible compared to the expansion of the
fluid, and the specific heat capacity at constant volume can be
approximated by that at constant pressure,
\begin{equation}
  c^v_{\partic} \approx c^p_{\partic}.
\end{equation}
We assume that the only contribution to the internal energy of the particles is
the thermal heat content,
\begin{align}
  \label{eq:e_p}
  e_\partic &= c^v_\partic T, \\
  \label{eq:e_p_approx}
           &\approx c^p_\partic T.
\end{align}
In assuming \cref{eq:e_p}, we have neglected e.g.\ the alignment energy of the
particles in the external magnetic field.
Using the definition of specific enthalpy~\eqref{eq:enthalpy_def} and
the approximation~\eqref{eq:e_p_approx}, we get an expression for the specific
enthalpy of the particles
\begin{align}
  \label{eq:h_p}
  h_\partic &\approx c^p_{\partic} T + \frac{p}{\rho_\partic},
\end{align}
in terms of $T$, $p$ and the constants $\rho_\partic$ and $c^p_\partic$.

\subsubsection{Ferrofluid}
\label{ssub:ferrofluid}

In our flow model~\eqref{eq:flow_steady_pmass}--\eqref{eq:flow_steady_h}, the
particle mass flux $\alpha_\partic \rho_\partic v$, base fluid mass flux
$\alpha_\basefl \rho_\basefl v$, and thus also the total mass flux $\rho v$, are
all constant along the pipe. Integration of \cref{eq:flow_steady_p} and
\cref{eq:flow_steady_h} provides the pressure $p$ and the
total enthalpy flux $\rho v h$ at every point along the pipe. The total enthalpy
flux is the sum of the particle and the base-fluid enthalpy fluxes,
\begin{equation}
  \label{eq:h_flux}
  \rho v h = \left( \alpha_\partic \rho_\partic v \right) h_\partic
    + \left( \alpha_\basefl \rho_\basefl v \right) h_\basefl.
\end{equation}
Since the fluid equations provide $p$, the
equilibrium calculation at the ferrofluid level is done by finding $T$ such
that \cref{eq:h_flux} is satisfied, where \cref{eq:basefluidtpflash}
and \cref{eq:h_p} are used to calculate $h_\partic$ and $h_\basefl$.

Having found the equilibrium temperature $T$ that satisfies \cref{eq:h_flux},
we use the constant $\rho_\partic$ and $\rho_\basefl$ from
\cref{eq:basefluidtpflash} to calculate the particle and base-fluid volume
fractions,
\begin{align}
  \alpha_\partic &= \frac{1}{1 + \frac{\left( \rho_\basefl \alpha_\basefl v
      \right)}{\rho_\basefl} \frac{\rho_\partic}{\left( \rho_\partic
        \alpha_\partic v \right)}}, \\
  \alpha_\basefl &= 1 - \alpha_\partic.
\end{align}
The total ferrofluid density is then found from
\begin{align}
  \rho &= \alpha_\partic \rho_\partic + \alpha_\basefl \rho_\basefl,
\end{align}
and the liquid and vapor volume fractions from
\begin{align}
  \alpha_\vap &= \alpha_\basefl
  \frac{\rho_\basefl - \rho_\liq}{\rho_\vap - \rho_\liq}, \\
  \alpha_l &= \alpha_\basefl - \alpha_\vap.
\end{align}

\subsection{Magnetization}
\label{sub:magnetization}

The magnetic force source term~\eqref{eq:force_term_mag} requires a model for
the ferroliquid magnetization $M_\ferrofl$, which we describe in this section.

The ferroliquid consists of
liquid base fluid and particles, but only the particles are
actually magnetized. Each particle has its own temperature-dependent
magnetization, we call it $M_\text{sat}$, that has an alignment with respect
to the external magnetic field. The ferroliquid magnetization may be found as
the average particle magnetization $M_\partic$ multiplied by the volume fraction
of particles in the ferroliquid,
\begin{align}
  M_\ferrofl = \frac{\alpha_\partic}{\alpha_\ferrofl} M_\partic.
  \label{eq:mag_f_from_mag_p}
\end{align}
Note that $M_\partic$ is the \emph{average} magnetization of the ensemble of
particles, and is not to be confused with $M_\text{sat}$, the magnetization of
a single particle alone. $M_\partic$ must be
smaller than or equal to $M_\text{sat}$, since the individual magnetic 
moments of each
particle are not completely aligned with each other, and thus cancel each other
out to some degree.

To what extent the magnetic moments of the particles are aligned is a
competition between the magnetic field and the temperature.
The magnetic field acts to align the moments, while the kinetic energy
associated with the temperature disrupts alignment. 
At very strong fields and/or very low temperatures,
which is called the saturated state,
all the particles are aligned, and thus the average particle magnetization
$M_\partic$ is equal to the magnetization of each particle $M_\text{sat}$. Due
to this correspondence, the magnetization of a single particle alone
is sometimes called the \emph{saturation magnetization} of the particle
ensemble.
It is also referred to as the \emph{spontaneous magnetization} of the
individual particles.

To model the effect of temperature-induced misalignment of the magnetic moments
of different particles, we employ the \emph{Langevin model},
\begin{equation}
  M_\partic(d_\partic,H,T) = M_\text{sat} (T)
  \mathcal{L}\left(\frac{V_\partic(d_\partic) M_\text{sat}(T) \mu_0 H}{k_B T}
  \right),
  \label{eq:mag_p_langevin}
\end{equation}
where $V_\partic$ (\si{\meter\cubed}) is the volume of a single particle, and
the \emph{Langevin function} is defined by
\begin{align}
  \mathcal{L} \left( x \right) \equiv \coth \left( x \right) - \frac{1}{x}.
  \label{eq:langevin}
\end{align}
This model can be derived from statistical mechanics by considering an ensemble
of non-interacting dipoles in an external magnetic field \cite{Andersen10}.

\subsubsection{Saturation magnetization}
\label{ssub:saturation_magnetization}

To account for the gradual loss of saturation magnetization as the
temperature approaches the \emph{Curie temperature} $T_\text{C}$,
$M_\text{sat}$ is modelled linearly as
\begin{equation}
  M_\text{sat}(T) =
    \begin{cases}
     M^\circ_{\text{sat}} \left(
       1 - \frac{T - T^\circ}{T_\text{C} - T^\circ}\right) & \text{if} \quad
     T<T_\text{C}, \\
       0 & \text{if} \quad T \geq T_\text{C},
    \end{cases}
  \label{eq:mag_sat}
\end{equation}
such that the particle magnetization is equal to a reference
magnetization $M^\circ_{\text{sat}}$ at the reference temperature $T^\circ$, and
zero when the temperature is above the Curie temperature of the particles.


\subsubsection{Particle size distribution}
\label{ssub:particle_size_distribution}

If all particles are of the same size, \cref{eq:mag_p_langevin} may be used
to represent the total average particle magnetization. We will refer to
this as the \emph{monodisperse model}.

To account for effects of particle size distribution on the magnetization, we
follow \textcite{Chantrell78} and integrate the particle magnetization
over a particular distribution of particle sizes,
\begin{equation}
  \label{eq:mag_p_log-normal}
  M_{\partic} (H,T) = \int_0^\infty
  f(x;\mu,\sigma) M_\partic(xd_\partic^{\text{med}},H,T)
  \dd x,
\end{equation}
where $M_\partic$ is given by \cref{eq:mag_p_langevin},
$d_\partic^{\text{med}}$ is the volume-weighted particle diameter median, $x =
d_\partic/d_\partic^{\text{med}}$ is the scaled particle diameter and $f$
is the log-normal probability density function,
\begin{equation}
  f(x;\mu,\sigma) =
  \frac{1}{x \sigma
    \sqrt{2\pi}} \exp \left(
    - \frac{(\ln(x) - \mu)^2}{2 \sigma^2} \right).
\end{equation}
This function describes the distribution of particle volume fractions, so that
$f(x;\mu,\sigma) \dd x$ is the volume fraction
of particles with scaled diameters between $x$ and $x + \dd x$. This means that
\begin{equation}
  \frac{1}{d_\partic^{\text{med}}}
  f(d_\partic/d_\partic^{\text{med}};\mu,\sigma) \dd d_\partic
\end{equation}
is the volume fraction of particles with diameters between $d_\partic$ and
$d_\partic + \dd d_\partic$.

The parameters $\mu$ and $\sigma$ describe the position and spread of the
particle size distribution. Specifically, they are related
to $d_\partic^{\text{med}}$ and the volume-weighted
diameter standard deviation $d_\partic^{\text{std}}$ by
\begin{align}
  \label{eq:mu}
  d_\partic^{\text{med}}/d_\partic^{\text{med}} &= \exp(\mu),\\
  d_\partic^{\text{std}}/d_\partic^{\text{med}} &= \sqrt{\left( \exp\left( \sigma^2
    \right) - 1 \right)
  \exp\left( 2 \mu + \sigma^2 \right)}.
\end{align}
Note that due to our choice of scaling, \cref{eq:mu} implies that we will
always have $\mu = 0$. 

We will refer to using \cref{eq:mag_p_log-normal} to calculate the total 
average particle magnetization as 
the \emph{log-normal model}.

\subsection{Nanofluid-modified transport properties}
\label{sub:transport_properties}
The addition of nanoparticles to a base fluid 
is known to affect transport properties such as thermal 
conductivity and viscosity~\cite{Kakacc09,Taylor13}.
If only the transport properties of the base fluid liquid is known in
advance, we have to estimate the effect of adding a certain amount of
nanoparticles. In the case of thermal conductivity, one may use 
the \emph{Maxwell model}~\cite{Maxwell1873,Sobti13} for the ferroliquid 
thermal conductivity,
\begin{equation}
  \lambda_\ferrofl = 
    \lambda_\liq \frac{\left(\frac{\lambda_\partic}{\lambda_\liq}+2\right)
            + 2\alpha_\partic \left(\frac{\lambda_\partic}{\lambda_\liq}-1\right)}%
          {\left(\frac{\lambda_\partic}{\lambda_\liq}+2\right)
            -  \alpha_\partic \left(\frac{\lambda_\partic}{\lambda_\liq}-1\right)},
  \label{eq:maxwell-thermal-conductivity}
\end{equation}
where 
$\lambda_k$ ($k \in \{\ferrofl,\liq,\partic\}$) is the conductivity of the ferroliquid, 
base liquid and particles, respectively. 
In the case of viscosity, one may use a 
higher-order extension~\cite{Sundar13, Brinkman52} of the 
\emph{Einstein model}~\cite{Einstein06}
for the ferroliquid dynamic viscosity, 
\begin{equation}
  \eta_\ferrofl =  \eta_\liq 
  \left(\frac{1}{1 - \alpha_\partic}\right)^{2.5},
  \label{eq:einstein-viscosity-higherorder}
\end{equation}
where $\eta_k$ ($k \in {\ferrofl,\liq}$) is the 
dynamic viscosity of the ferroliquid and base fluid liquid, respectively.

However, if the transport properties of the ferroliquid have been
measured directly, it may be preferable to use those constant values,
instead of using the above models to modify the base fluid properties.


\subsection{Solenoid magnetic field}
\label{sub:solenoid_magnetic_field}

In order to calculate the magnetic force term~\eqref{eq:force_term_mag},
we need the derivative of the magnetic $H$-field with respect to
axial position $x$. We also need the magnetization, which in turn
requires the magnetic $H$-field itself.

The magnetic field along the axis of an empty solenoid of finite length and
width is \cite{Jiles98}
\begin{align}
  \label{eq:H_solenoid_finite_empty}
  H_x &= \frac{n I}{2(R_2-R_1)}
    \left[(x-x_1) \xi_1 - (x-x_2) \xi_2 \right],
\end{align}
with the derivative
\begin{multline}
  \label{eq:dHdx_solenoid_finite_empty}
  \frac{\dd H_{x}}{\dd x}
  = \frac{n I}{2(R_2-R_1)} \Bigg[
    \xi_1 - \xi_2 \\
    + \left(\frac{(x-x_1)^2}{(R_2 + \xi_{21})\xi_{21}}
      - \frac{(x-x_1)^2}{(R_1 + \xi_{11})\xi_{11}}\right) \\
    - \left(\frac{(x-x_2)^2}{(R_2 + \xi_{22})\xi_{22}}
      - \frac{(x-x_2)^2}{(R_1 + \xi_{12})\xi_{12}}\right) \Bigg],
\end{multline}
where
\begin{align}
  \label{eq:H_solenoid_xidef}
  \xi_{ij} &= \sqrt{R_i^2 + (x-x_j)^2},\\
  \xi_j &= \ln\left(\frac{R_2+\xi_{2j}}
                       {R_1+\xi_{1j}}\right).
\end{align}
In the above, 
$n$~(\si{\meter\tothe{-1}}) is the number of wire windings per axial
length of the solenoid, $I$~(\si{\ampere}) is the wire current,
$R_1$~(\si{\meter}) and $R_2$~(\si{\meter}) are the inner and outer radii of the
solenoid, respectively, $x_1$~(\si{\meter})
is the position of the left end, and $x_2$~(\si{\meter}) is the position of the
right end. Here we have assumed that the current is evenly distributed across
the cylindrical sheet. The $x$-axis is taken to be along the central axis of
the solenoid, directed such that the current runs clockwise when looking in the
positive direction. The expression may be found by integrating Biot--Savart's
law in both the axial and the radial direction.
%
The derivative peaks at \emph{approximately} $x_1$ and $x_2$.
When the solenoid is significantly wider than the pipe, the axial field above
may be used as an approximation to the field over the entire pipe cross
section. 

Note that in this case, the solenoid is not in fact empty, but has a
magnetizable ferrofluid core. However, here we still use 
\cref{eq:H_solenoid_finite_empty,eq:dHdx_solenoid_finite_empty} to
calculate the axial $H$-field, due to the following argument.
Inside a finite solenoid, we know from Ampère's law that 
    the axial integral of the axial $H$-field is independent of the 
    core material ($\int_{-\infty}^{\infty}H_x \mathrm{d}x = nI(x_2-x_1)$).
    We also know that $H_x$ has the same sign along the entire solenoid axis.
    This indicates that the presence of 
    a weakly magnetizable core will only give a slight redistribution 
    of $H_x(x)$, without changing its integrated magnitude.
Finite Element calculations using the FEMM~\cite{femm} software
    confirmed this for the case of 
    magnetizable pipe contents with the susceptibility of the 
    ferrofluid used in this study ($\chi \approx 0.1$). 
Note that the $B$-field is in fact affected significantly by the presence
of the ferrofluid. However, the $B$-field does not enter our model equations, 
since the Kelvin force in \cref{eq:force_kelvin_1D} only depends on the
$H$-field.

The above is only a calculation of the $H$-field exactly along the central
solenoid axis. Using this as the $H$-field in the one-dimensional pipeline
model will be a good approximation if the pipeline diameter is much smaller
than the solenoid diameter, which is the case in this work.

\section{Numerical methods}
\label{sec:numerics}

\subsection{Flow equations}
\label{sub:ode_integration}
The fluid equations \eqref{eq:flow_steady_p}--\eqref{eq:flow_steady_h} are
integrated using \verb|SciPy|'s \cite{Scipy} bindings to \verb|LSODA| from the
\verb|ODEPACK| library \cite{Hindmarsh83}. We demand a maximum relative error
of \SI{e-12}{}. The terms involving $\dd v/ \dd x$ on the right-hand sides 
of~\eqref{eq:flow_steady_p} and~\eqref{eq:flow_steady_h} are not calculated, 
since they were found to be negligible compared to the other terms. This 
simplifies the integration, as the right-hand sides then only depend on
the local state, and no derivatives (except $\dd H/\dd x$, which is
static). Calculating the right-hand sides
of~\eqref{eq:flow_steady_p} and~\eqref{eq:flow_steady_h} then involves
calculating the new thermodynamic equilibrium state of the ferrofluid
based on the current primary flow variables, and then 
calculating each source term based on that.

\subsection{Thermodynamic equilibrium}
\label{sub:thermodynamic_equilibrium}

\subsubsection{Base fluid}
The set of non-linear equations solved to find the equilibrium state, 
and bubble point, of the base
fluid are solved using an in-house implementation of the methods described in
\textcite{Michelsen07}.

\subsubsection{Ferrofluid}
To find the equilibrium state of the ferrofluid, \cref{eq:h_flux} is solved for
the temperature $T$. This is done using the secant method implemented in the
\verb|SciPy| library \cite{Scipy}. We demand a maximum relative error of
\SI{e-9}{}.

\subsection{Log-normal particle distribution}
\label{sub:lognormal_particle_distribution}
The average particle magnetization is modelled by the integral
\eqref{eq:mag_p_log-normal}. It is evaluated numerically using
\verb|scipy.integrate.quad| from \verb|SciPy|~\cite{Scipy}, which uses the
method \verb|QAGS| from the Fortran library \verb|QUADPACK|~\cite{Piessens83}.
We demand a maximum relative error of \num{e-8}.

\section{Validation}
\label{sec:validation}
The model was validated against
experiments presented in~\textcite{Iwamoto11}, with
focus on correctly predicting the pressure increase achieved 
from the thermomagnetic pump. 
In \cref{sub:Representing the ferrofluid},
\ref{sub:The experimental rig}
and~\ref{sub:heat transfer coefficient}, we show how the 
ferrofluid, the experimental setup, and the wall heat transfer
are represented in this model. The results of the validation are then shown 
in \cref{sub:results}.

\subsection{Representing the ferrofluid}
\label{sub:Representing the ferrofluid}
The experiments in \cite{Iwamoto11} were performed on a mixture of
a commercial ferrofluid (``TS50K'') and n-Hexane.
In terms of the model, the ferrofluid is represented in two parts, 
the base fluid and the particles. 
The base fluid is represented by 
a thermodynamic model (Sec.~\ref{ssub:base_fluid}), 
which can supply predictions for
most needed properties, including the possible appearance of vapor.
The solid particle phase is represented by a set of constant thermodynamic
properties (Sec.~\ref{ssub:particles}), 
and a magnetization model fitted to experimental magnetization data
(Sec.~\ref{sub:magnetization}).
The thermodynamic properties of the total fluid are then based on the models 
for the base fluid and the particles, and the local particle volume
fraction (Sec.~\ref{ssub:ferrofluid}).

Due to the imperfect predictions of the equation of state, choosing 
parameters to best represent the given ferrofluid is not straightforward.
How it was done, and the choices made along the way, is described in this
section.

\subsubsection{Fluid properties}
\label{ssub:fluid_properties}
The properties of the fluids in~\cite{Iwamoto11} were measured in a state
below the boiling point, and thus in a two-phase liquid--particle state. 
Some relations are needed in order to derive properties additional to
the ones measured directly. The volume fraction of 
particles can be calculated from the relation
\begin{equation}
  \alpha_\partic = w_\partic \frac{\rho}{\rho_\partic}, 
  \label{eq:w_to_alpha}
\end{equation}
where $w_k$~(--) is the mass fraction of component $k$.

\paragraph{TS50K:} 
~\\
The commercial ferrofluid TS50K is a mixture of 
50 wt\% kerosene, 50 wt\% MnZn ferrite nanoparticles. 
Its relevant known properties are summarized in \cref{tab:ts50k}.
This is not the fluid used in the main experiments in~\cite{Iwamoto11}, 
but it is the fluid for which the magnetization response had been measured.

\begin{table}[htbp]
  \centering
  \caption{Properties of TS50K (kerosene + MnZn ferrite nanoparticles) at
    \SI{293}{\kelvin} and \SI{1.013}{\bar}.}
  \begin{tabular}{lcrc}
    \toprule
    Description &  & Value & Ref.\\
    \midrule
    Density & $\rho$ & \SI{1401}{\kilogram\per\cubic\meter}  &
      \cite{Iwamoto11}\\
    Specific heat cap.\ & $c^p$
      & \SI{1387}{\joule\per\kilogram\per\kelvin} & 
      \cite{Iwamoto11}\\
      Particle density & $\rho_\partic$ & \SI{5000}{\kilogram\per\cubic\meter} &
      \cite{Iwamoto20140417}\\
    Particle mass frac.\ & $w_\partic$ & 0.5 & \cite{Iwamoto20140417}\\
    Particle vol.\ frac.\ & $\alpha_\partic$ & 0.140 & 
      \eqref{eq:w_to_alpha}\\
    \bottomrule
  \end{tabular}
  \label{tab:ts50k}
\end{table}

\paragraph{TS50K with n-Hexane:} 
~\\
The main experiments in~\cite{Iwamoto11} were performed on a 
\emph{binary TSMF}, which was
TS50K mixed with hexane at a ratio of 80/20 wt\%. 
This means that the total fluid was a 
40/40/20 wt\% mixture of particles, kerosene and hexane, respectively.
Due to the addition of hexane, the particle volume fraction and
liquid density must be recalculated.
The relevant known properties of this fluid are summarized in 
\cref{tab:ts50khexane}.

\begin{table}[htbp]
  \centering
  \caption{Properties of TS50K with hexane at \SI{293}{\kelvin}
    and \SI{101.3}{\kilo\pascal}.}
  \begin{tabular}{lcrc}
    \toprule
    Description & & Value & Ref.\\
    \midrule
    Density & $\rho$ & \SI{1143}{\kilogram\per\cubic\meter} & 
      \cite{Iwamoto11}\\
    Particle density & $\rho_\partic$ & \SI{5000}{\kilogram\per\cubic\meter} &
      \cite{Iwamoto20140417}\\
    Liquid density & $\rho_\liq$ & \SI{754.8}{\kilogram\per\cubic\meter} &
      \eqref{eq:total_rho_from_sum}\\
    Viscosity & $\eta$ & \SI{2.35}{\milli\pascal\second} &
      \cite{Iwamoto11}\\
    Specific heat cap.\ & $c^p$
      & \SI{1564}{\joule\per\kilogram\per\kelvin} & 
      \cite{Iwamoto11}\\
    Thermal cond.\ & $\lambda$ & \SI{0.16}{\watt\per\meter\per\kelvin} &
      \cite{Iwamoto11}\\
    Bubble temp.\ & $T_\mathrm{bub}$ & \SI{358}{\kelvin} &
      \cite{Iwamoto11}\\
    Particle mass frac.\ & $w_\partic$ & 0.4 &
      \cite{Iwamoto11}\\
    Particle vol.\ frac.\ & $\alpha_\partic$ & 0.0914 &
      \eqref{eq:w_to_alpha}\\
    \bottomrule
  \end{tabular}
  \label{tab:ts50khexane}
\end{table}

The kerosene in TS50K is in fact a mixture of over 
20 components~\cite{Iwamoto20140417}, and it is not feasible to 
represent it completely in the equation of state. 
In this work, we model the base fluid as a binary mixture of 
undecane (\ce{C11H24}) and hexane (\ce{C6H14}), and set the relative amounts
such as to fit the measured properties of the actual base fluid
as well as possible.

The known properties of the base fluid in \cref{tab:ts50khexane} 
are the liquid density and the bubble temperature. 
It is not necessarily possible to satisfy both exactly.
A fitting to data at the given temperature and pressure 
led to a mixture of 65\% mass fraction undecane.
This gives a near perfect fit for the 
bubble temperature, and a liquid density 
of $\SI{722.4}{\kilogram\per\cubic\meter}$, which is a deviation of
about $4\%$ from the actual value.
At this point, the equation of state predicts a specific heat capacity 
of $\SI{2201}{\joule\per\kilogram\per\kelvin}$ for the liquid.


The primary goal of the validation is the correct prediction of the pressure 
jump achieved by the thermomagnetic pump. The magnetic force scales 
with $\alpha_\partic$ and the magnetization. The magnetization scales
with the temperature, whose change depends on the volumetric heat 
capacity ($\rho c^p$). We therefore make the choice that the model fluid
should reproduce the exact values for $\alpha_\partic$, $\rho$ 
and $c^p$ for the actual fluid, which are given in \cref{tab:ts50khexane}. 
Due to the imperfect liquid density prediction,
obtaining the actual $\alpha_\partic$ and $\rho$ 
demands a change to
$\rho_\partic = \SI{5323}{\kilogram\per\meter\cubed}$ and
$w_\partic = 0.426$.

The specific heat capacity of the particle material is not known exactly.
One may find sources for the heat capacity of some MnZn 
ferrites (in the area of \SIrange{500}{1500}{\joule\per\kilogram\per\kelvin}), 
but they will vary considerably, 
and depend on the exact composition. The fact that the substance is present
as nanoparticles instead of in bulk form will also affect
physical properties.
However, since 
the combined specific heat capacity is a mass-weighted sum of the 
specific heat capacities of the phases,
obtaining the actual $c^p$ demands that we now set 
$c^p_{\partic} = \SI{704.5}{\joule\per\kilogram\per\kelvin}$.

As implied by \cref{eq:flow_steady_pmass}, the local particle volume
fraction $\alpha_\partic$ in the model will change as the velocity
changes. The input parameter is thus the inlet volume fraction,
$\alpha_\partic^\circ$, which we set equal to the overall 
volume fraction in \cref{tab:ts50khexane}.

In this case, the transport properties of the ferroliquid phase are
given in~\cite{Iwamoto11}. The changes in ferroliquid
conductivity and viscosity due to the presence of nanoparticles are
thus not modelled, and these constant values are instead applied to
the ferroliquid phase.  The model also requires transport properties
for the vapor phase, in the flow model if $\alpha_\vap>0$, and in some
heat transfer coefficient correlations. Here we use constant transport
properties for the vapor.  Since hexane is the lightest component, it
will be the main component of the gas phase, hence we use values for
hexane. Some heat transfer correlations also depend on surface
tension. We weight the surface tension of the two components with
liquid mole fraction \cite{Eberhart66}. The values for vapor viscosity
and surface tension used in the model were retrieved from the NIST
Chemistry Webbook \cite{NISTwebbook} and are shown in
\cref{tab:model_parameters}.



%

%



\subsubsection{Magnetization response}
\label{ssub:nanoparticles}

Measured data of the magnetization response of TS50K,
before adding hexane, was available~\cite{Iwamoto20140417}. 
The measurements consist of two curves: Varying $H$
at constant $T^\circ=\SI{300.1}{\kelvin}$, and 
varying $T$ at constant $H^\circ=\SI{796.0}{\kilo\ampere\per\meter}$.

\begin{figure}
        \centering
        \begin{subfigure}[b]{0.48\textwidth}
                \centering
                \includegraphics[width=\textwidth]{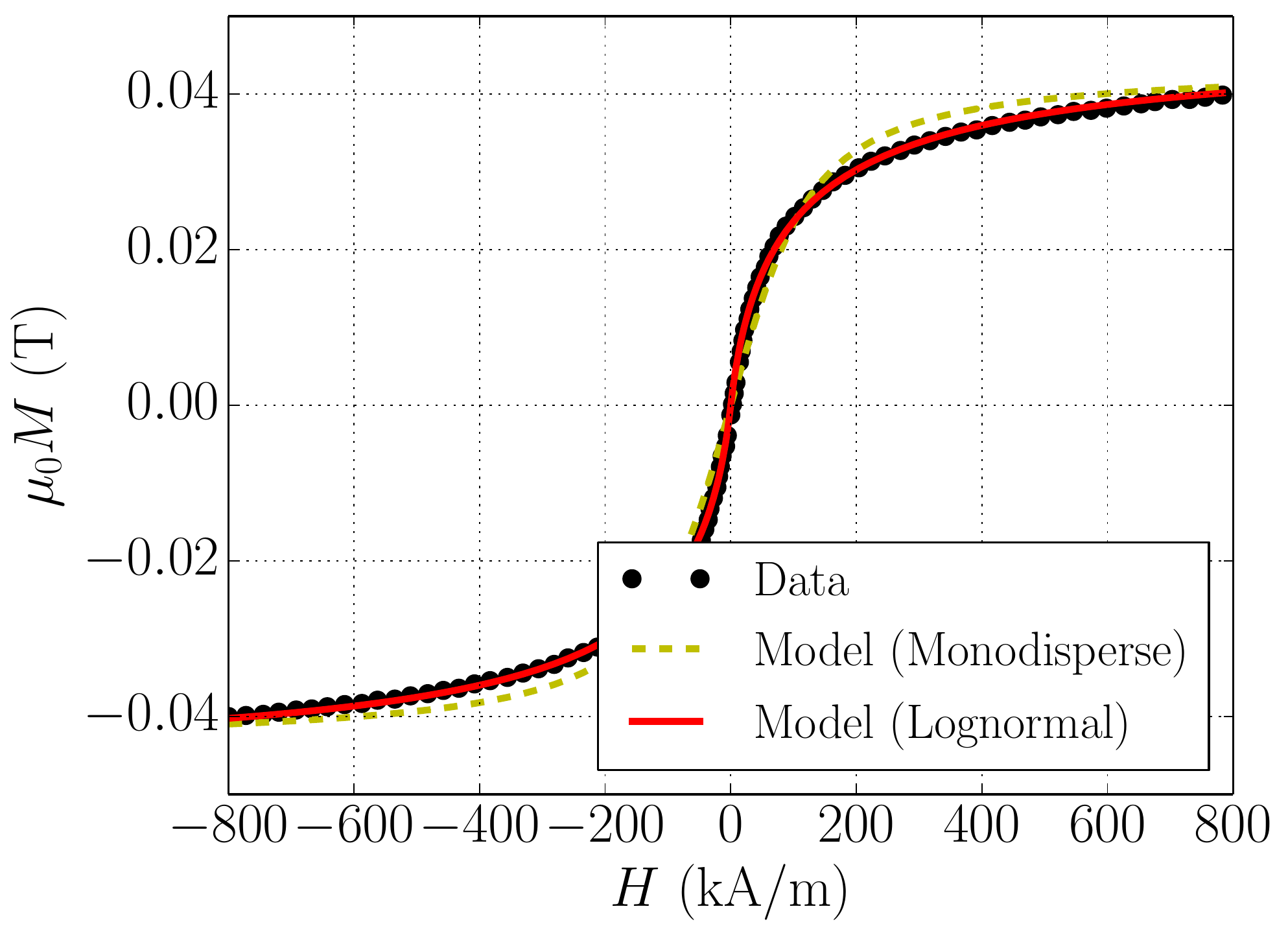}
                \caption{}
                \label{fig:fitted_mag_MH}
        \end{subfigure}%
        \\
        \begin{subfigure}[b]{0.48\textwidth}
                \centering
                \includegraphics[width=\textwidth]{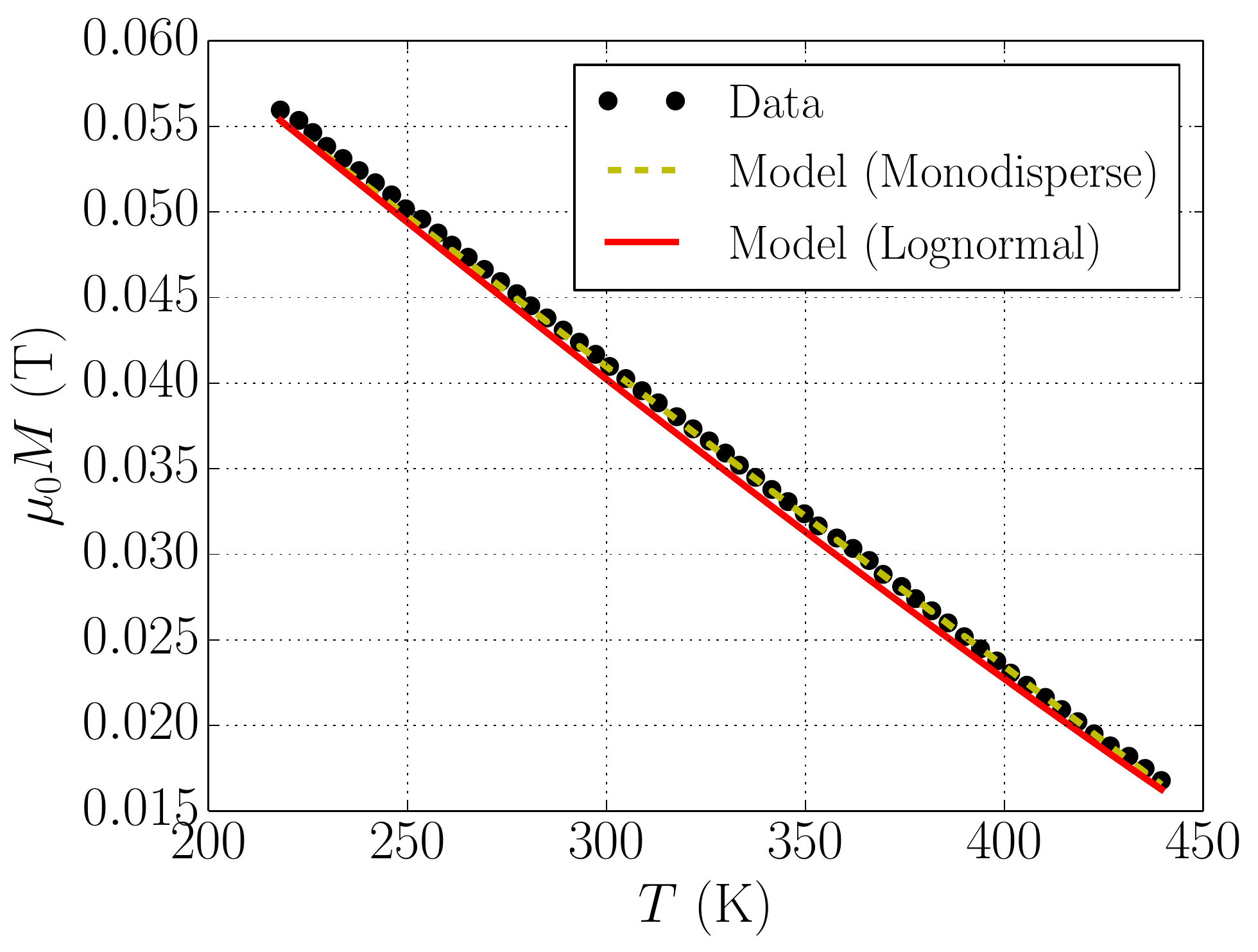}
                \caption{}
                \label{fig:fitted_mag_MT}
        \end{subfigure}
        \caption{Fitting of models to magnetization data with (a)
          varying $H$ at constant $T=\SI{300.1}{\kelvin}$, and 
          (b) varying $T$ at constant $H=\SI{796.0}{\kilo\ampere\per\meter}$.}
        \label{fig:fitted_mag}
\end{figure}

This data was used to fit the parameters of the models for
average particle magnetization, described in 
\cref{sub:magnetization}. 
This is three parameters for the monodisperse model
($d_\partic$, $M^\circ_{\text{sat}}$, $T_\text{C}$)
and four parameters for the log-normal distribution model
($d_\partic^{\text{med}}$, $d_\partic^{\text{std}}$, 
$M^\circ_{\text{sat}}$, $T_\text{C}$).
The results of the curve fitting procedures 
are shown in \cref{fig:fitted_mag}, for both models. 
As seen, the log-normal distribution model has a very good fit
to the data, and is used in the remaining validation. 
The offset in \cref{fig:fitted_mag_MT} is due to
an inconsistency in the data, where the magnetization at the common point 
$H^\circ=\SI{796.0}{\kilo\ampere\per\meter}$
$T^\circ=\SI{300.1}{\kelvin}$ is slightly higher in 
\cref{fig:fitted_mag_MT} compared to 
\cref{fig:fitted_mag_MH}.

\begin{figure}
        \centering
        \includegraphics[width=0.4\textwidth]{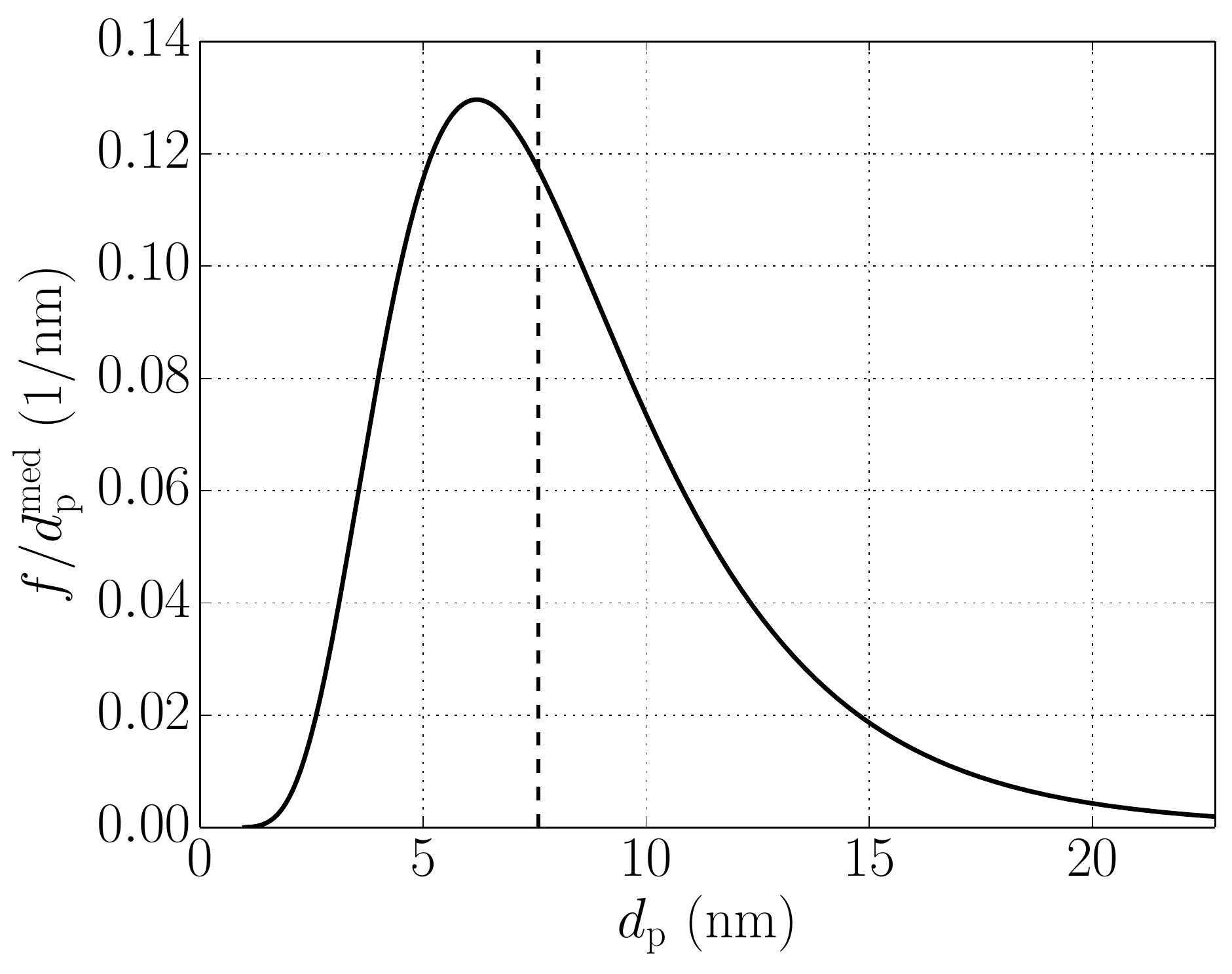}
        \caption{The log-normal diameter distribution fitted to 
          magnetization data, with the median shown as a dashed line.
          The distribution represents how the total volume of particles
          is distributed across different diameters.}
        \label{fig:lognorm_distr}
\end{figure}

These models may then also be used to describe the magnetization
response in TS50K with hexane, since the particles are the same. 
The total magnetization will be lower though, since $\alpha_\partic$
is lower. The parameters of the best fit are shown in 
\cref{tab:model_parameters}, and the corresponding size
distribution is shown in \cref{fig:lognorm_distr}.

\subsubsection{Model ferrofluid parameters}
\label{ssub:model_ferrofluid_parameters}
A summary of all the adjustable parameters in the model
used in this work is shown in \cref{tab:model_parameters}.
The equation of state for the base fluid involves additional
parameters, but they are not listed here, as they are 
not specific for this model or case, but part of a
generalized thermodynamic library.

\begin{table}[htbp]
  \centering
  \caption{Summary of parameters defining the ferrofluid in this model. 
      Parameters marked with N/A are relevant for the model in general, but
      not in the specific validation cases in this work.}
  \begin{threeparttable}[b]
  \begin{tabular}{lcr}
    \toprule
    Description &  & Value \\
    \midrule
    \textbf{Ferroliquid}\\
    Viscosity & $\eta_\ferrofl$ 
                          & \SI{2.35e-3}{\pascal\second} \\
    Thermal cond. & $\lambda_\ferrofl$ 
                          & \SI{0.16}{\watt\per\meter\per\kelvin} \\
    \textbf{Base fluid}\\
    Vapor viscosity & $\eta_\vap$
                          & \SI{7.98e-6}{\pascal\second}\\
    Vapor thermal cond. & $\lambda_\vap$
                          & \SI{0.0188}{\watt\per\meter\per\kelvin} \\
    Components & & \ce{C11H24}, \ce{C6H14}\\
    Mass composition & & 65\%/35\% \\
    Hexane surf. tens. & $\sigma_1$ & \SI{1.17e-2}{\newton\per\meter} \\
    Undecane surf. tens.\tnote{1}
    & $\sigma_2$ & \SI{1.88e-2}{\newton\per\meter} \\
    \textbf{Particles}\\
    Density & $\rho_\partic$ & \SI{5323}{\kilogram\per\cubic\meter} \\
    Spec.\ heat cap.\ & $c^p_{\partic}$ &  
      \SI{704.5}{\joule\per\kilogram\per\kelvin}\\
    Thermal cond.\ & $\lambda_\partic$ 
                                  & (N/A) \\
    Inlet volume fraction  & $\alpha_\partic^\circ$ & 0.0914 \\
    Mean diameter & $d_\partic^{\text{med}}$ & 
         \SI{7.58}{\nano\meter}  \\
    Diameter std.dev. & $d_\partic^{\text{std}}$ &  
        \SI{3.96}{\nano\meter} \\
    Sat.\ mag.\ at $T^\circ$
          & $M^\circ_{\text{sat}}$ &  
        \SI{265}{\kilo\ampere\per\meter}\\
    Curie temperature & $T_\text{C}$ &  \SI{576}{\kelvin} \\
    \bottomrule
  \end{tabular}
  \begin{tablenotes}
    \item[1] {\footnotesize Average of values for decane and dodecane}
  \end{tablenotes}
  \end{threeparttable}
  \label{tab:model_parameters}
\end{table}

\subsection{The experimental rig}
\label{sub:The experimental rig}
\begin{figure}[htpb]
  \centering

  \includegraphics[width=0.46\textwidth]{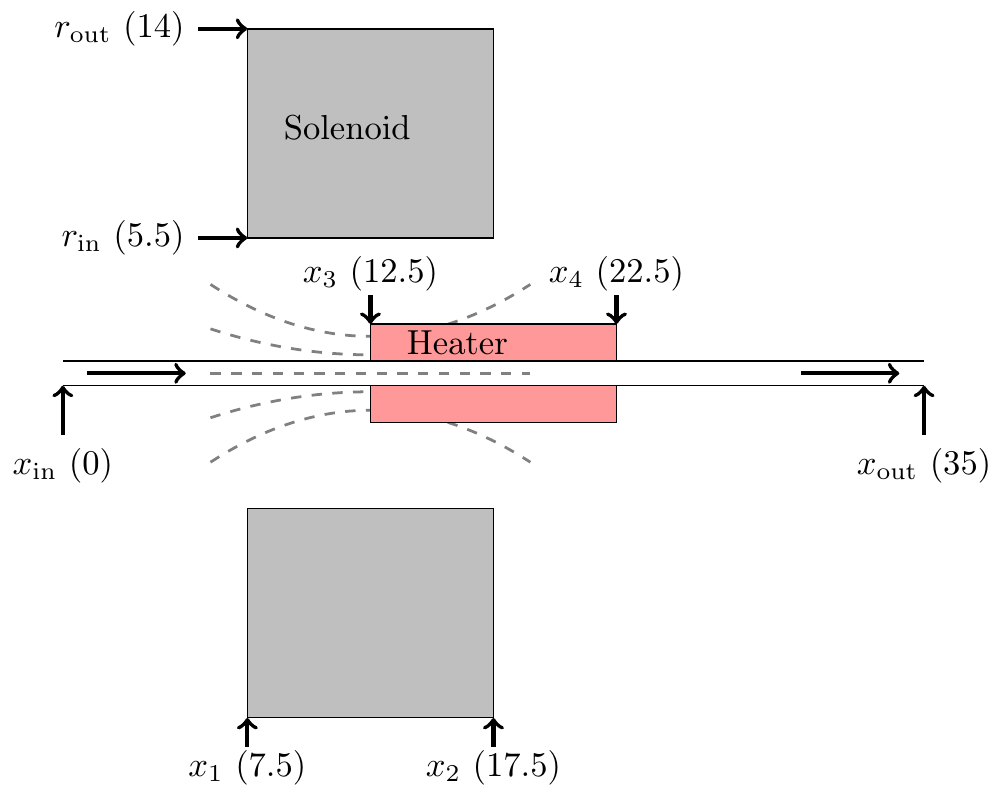}

  \caption{A cross section of the experimental rig in~\cite{Iwamoto11},
    to scale in both the axial and the radial direction. The rig geometry 
    is rotationally symmetric around the pipe axis.
    With the origin at the left and center of the pipe,
    numbers show positions in \si{\centi\meter}, 
    with vertical and horizontal arrows giving axial and radial positions,
    respectively. Dashed lines illustrate the magnetic field lines giving
    magnetic pumping force.
    The direction of gravity with respect to this figure
    varies with the tilt of the rig. In the horizonal configuration, 
    gravity points perpendicular to the flow direction. 
    In the vertical configuration, gravity points opposite to 
    the flow direction.}
  \label{fig:iwamoto_rig}
\end{figure}

The experimental rig of~\textcite{Iwamoto11} consisted of
a thermomagnetic pump (solenoid and heater) wrapped around a
\SI{1}{\centi\meter} pipe,
all of which could be tilted. In each case, the pressure difference across
specified inlet and outlet points around the thermomagnetic pump was 
measured.
A control system elsewhere in the loop kept the inlet conditions
at approximately $p_\mathrm{in}=\SI{1.163}{\bar}$,
$T_\mathrm{in}=\SI{307}{\kelvin}$ and 
$v_\mathrm{in}=\SI{2.0}{\centi\meter\per\second}$.  
The exact geometry of the setup is shown in 
\cref{fig:iwamoto_rig},
drawn from~\cite{Iwamoto11} 
and additional information supplied by its authors~\cite{Iwamoto20140925}.

The geometry of the solenoid is shown to scale. 
It consists of about 850 turns of copper wire.
With the given geometry, fitting these turns requires
a wire diameter of about $\SI{3.4}{\milli\meter}$, 
giving a solenoid resistance of about $\SI{1}{\ohm}$.
Obtaining $H_\mathrm{max} = \SI{100}{\kilo\ampere\per\meter}$ 
with this solenoid requires a current
of $\SI{25}{\ampere}$, giving a voltage-drop of 
$\SI{25}{\volt}$ and thus a power dissipation of about
$\SI{600}{\watt}$.

The heater is modelled as a constant pipe wall temperature, while the
pipe is approximated as perfectly insulated elsewhere. The heater temperature is
specified in relation to $T_\mathrm{in}$, as a difference $\Delta T$.

\subsection{Heat transfer coefficient}
\label{sub:heat transfer coefficient}
An important model parameter is the heat transfer coefficient (HTC), 
which is needed when calculating the wall heat transfer term
described in~\cref{ssub:heat_transfer}. This may either be set constant,
or represented by one of many available correlations. 

For the cases with $\Delta T = \SI{70}{\kelvin}$, the heater wall temperature
is above the bubble temperature of the fluid, and boiling heat transfer
occurs. This is either subcooled or saturated boiling, depending on whether
the local average fluid temperature is below or above the bubble temperature.
Boiling heat transfer correlations have mostly been made for either saturated
or subcooled boiling. 
A single validation case presented here may include both regimes, as the
temperature rises along the pipe.
For this reason, it is preferable to use correlations which extend to both
regimes.

In this work, we attempt to use two different correlations where boiling
occurs:
\textcite{Gungor86}, which claims to give reasonable results in 
both regimes, and
\textcite{Chen66}, which was originally developed for saturated
boiling, but is recommended for use in subcooled boiling 
in~\cite{Gungor87}.
The equations do not explicitly account for the presence of nanoparticles 
in the liquid, but when used here we supply the ferroliquid transport 
properties in place of liquid transport properties. This should be
sufficiently accurate under the assumption that the effects of boiling on heat transfer
dominate over the effects of the nanoparticles.



\subsection{Results}
\label{sub:results}

Simulations were run with the ferrofluid parameters shown in
\cref{sub:Representing the ferrofluid} and the 
rig set-up shown in \cref{sub:The experimental rig}.
Here the focus is the effect of the magnetic field strength on
the pressure jump from the inlet to the outlet, $\Delta p$,
which represents the thermomagnetic pumping effect. 
The quantity $\Delta p$ is thus zero at zero field in each case, by
definition. The quantity is plotted against $H_\mathrm{max}$,
which is the maximum field in the solenoid. 

\begin{figure}
        \centering
        \includegraphics[width=0.46\textwidth]{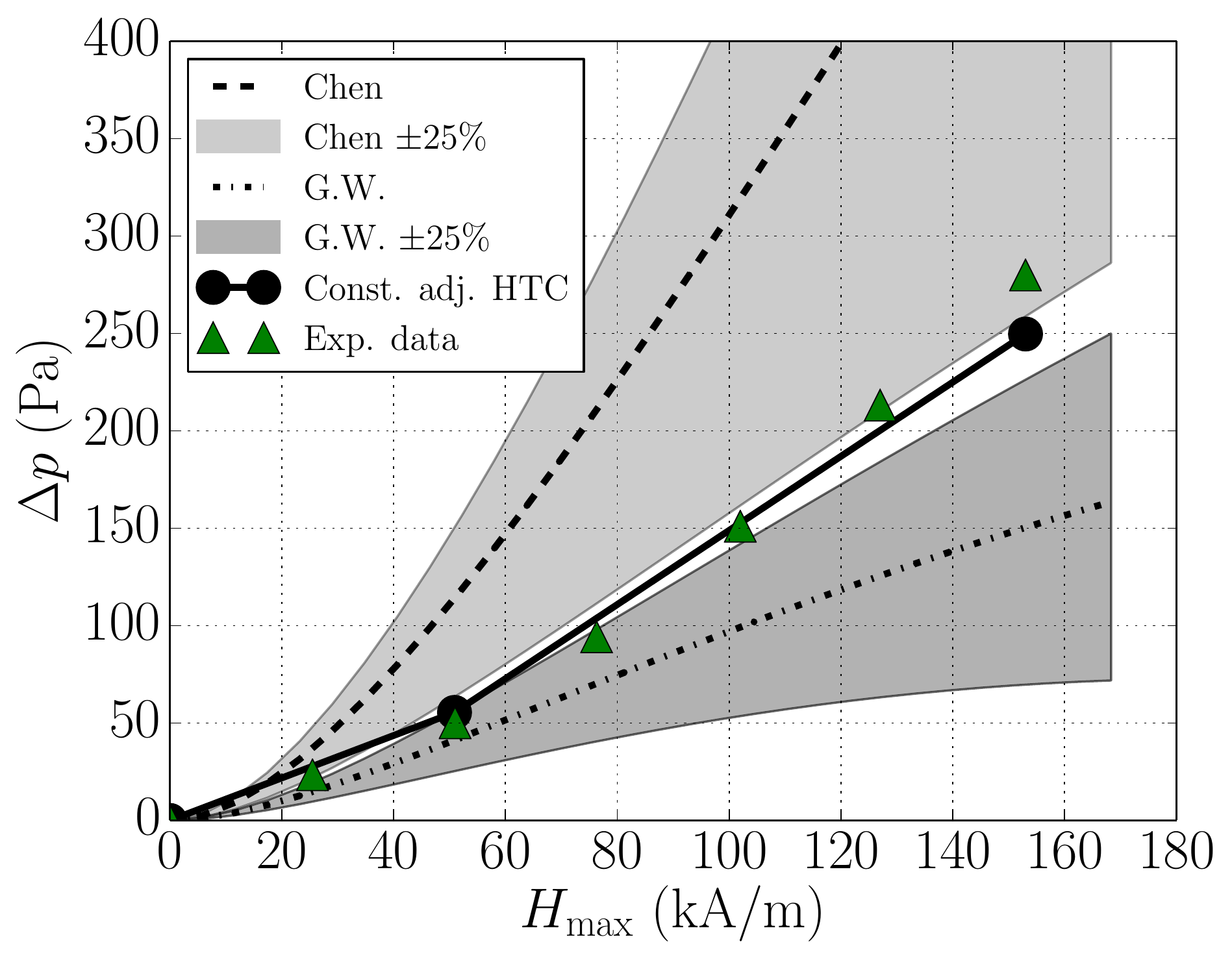}
        \caption{Results for the horizontal case with 
                 $\Delta T=\SI{70}{\kelvin}$, compared with experiments
                 from~\cite{Iwamoto11}. 
                 }
        \label{fig:deltap_vs_Hmax_0deg_70K}
\end{figure}

\begin{figure}
        \centering
        \includegraphics[width=0.46\textwidth]{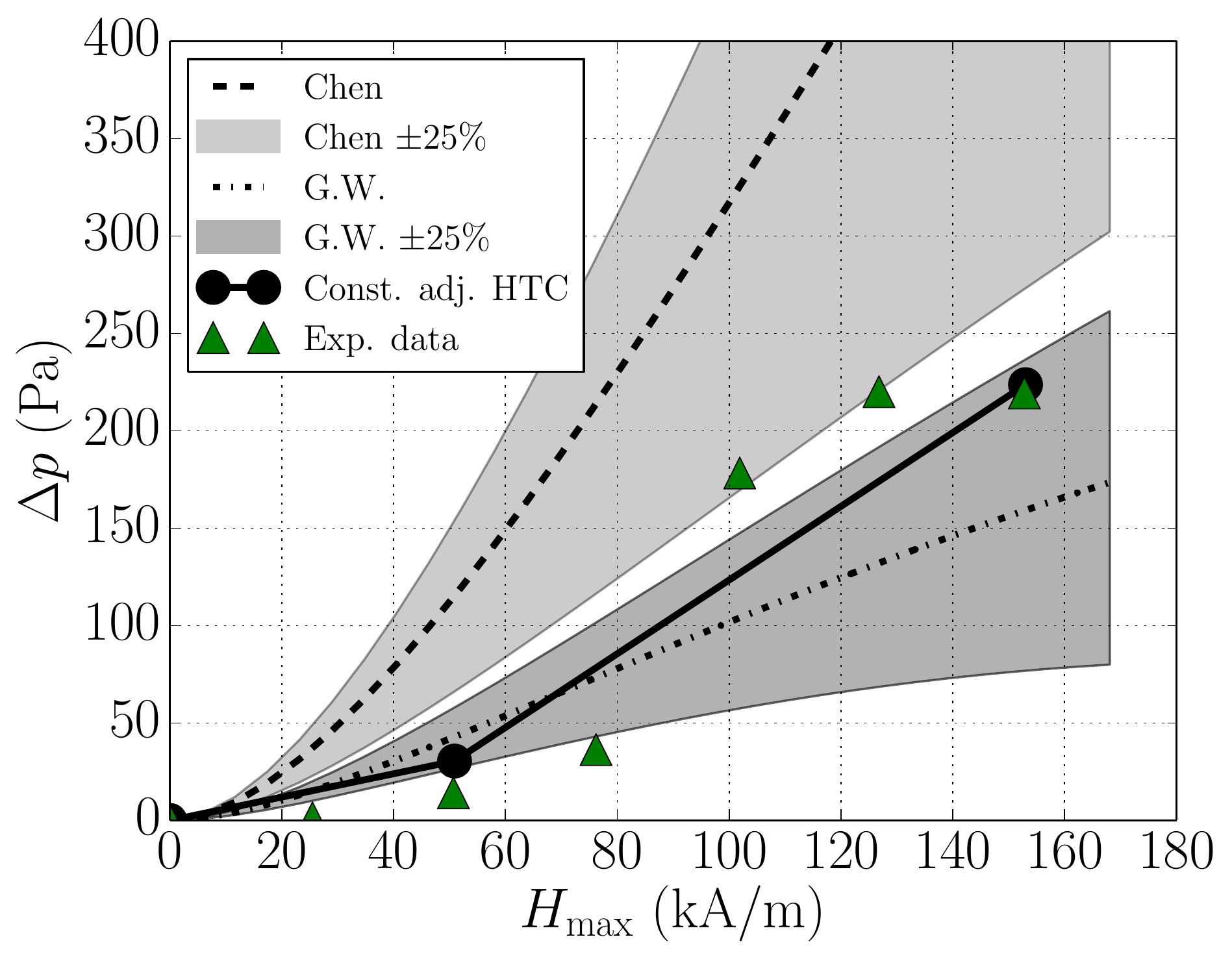}
        \caption{Results for the vertical case with 
                 $\Delta T=\SI{70}{\kelvin}$, compared with experiments
                 from~\cite{Iwamoto11}. 
                 }
        \label{fig:deltap_vs_Hmax_90deg_70K}
\end{figure}

For the case of $\Delta T=\SI{70}{\kelvin}$, data sets are available
in~\cite{Iwamoto11} for both the horizontal and vertical orientation.
Here boiling heat transfer occurs, and simulations were run with the
two boiling HTC correlations in
\cref{sub:heat transfer coefficient}. To demonstrate the sensitivity of
the predictions to this, bands of prediction given a 25\% 
uncertainty in the HTC are also shown. This is a common uncertainty
estimate, but the actual uncertainty can be much larger in certain
parameter areas~\cite{Gungor86}. 
For two values of $H_\mathrm{max}$, the outlet temperatures are 
available. In these cases, one may also adjust a constant HTC to achieve
the correct outlet temperature in the model. Here the HTC in the reference
zero-field point is assumed equal to the HTC in the point of lowest field.
All these results are plotted together with the experimental data
from~\cite{Iwamoto11} in
\cref{fig:deltap_vs_Hmax_0deg_70K}
and 
\ref{fig:deltap_vs_Hmax_90deg_70K}
for the horizontal and vertical case, respectively.

\begin{figure}
        \centering
        \includegraphics[width=0.46\textwidth]{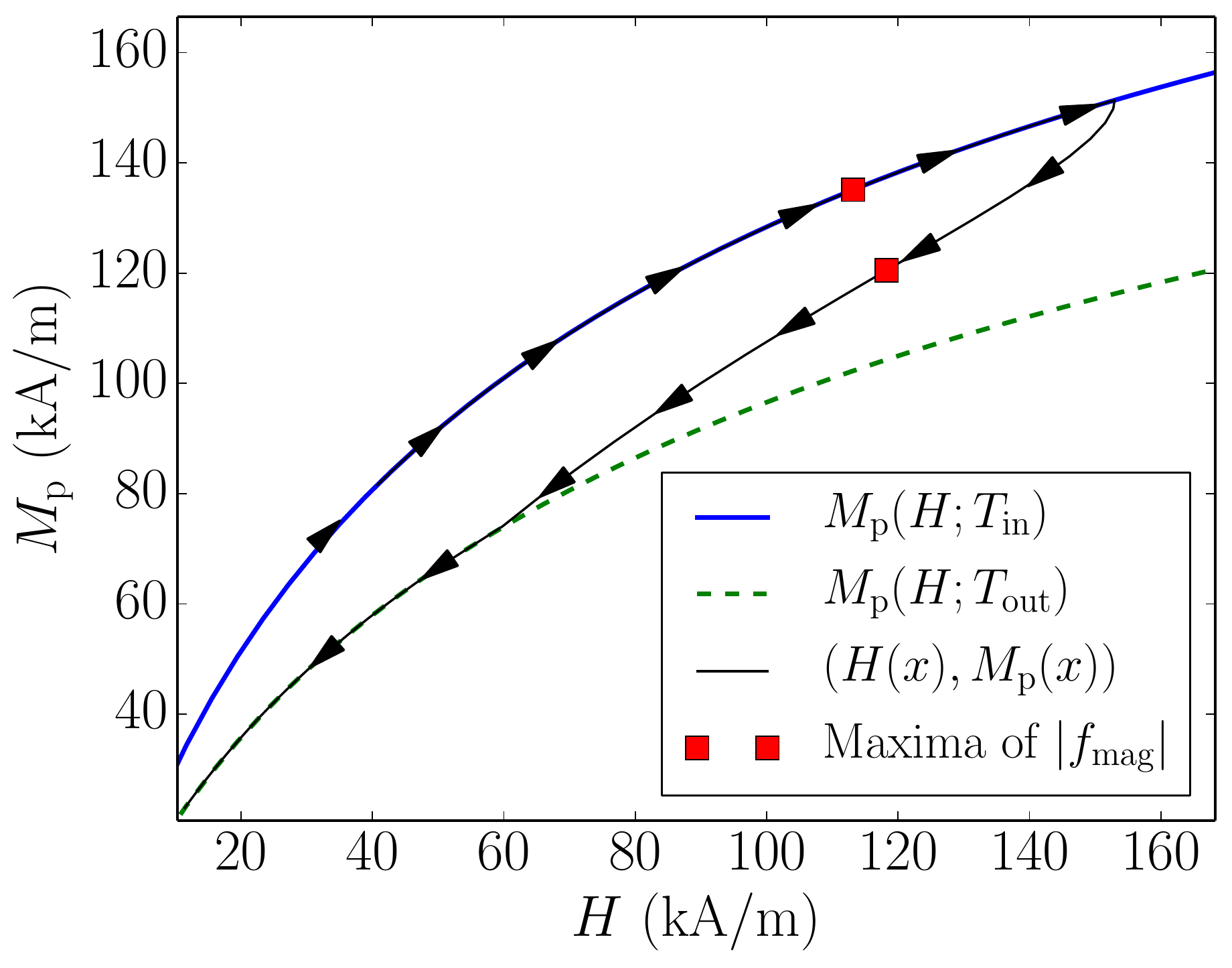}
        \caption{The particle magnetization curves at the inlet
          and outlet temperatures, and the path taken along
          the pipe,
          for a case with $\Delta
          T=\SI{70}{\kelvin}$ and $H_\mathrm{max} =
          \SI{153}{\kilo\ampere\per\meter}$.}
        \label{fig:mag_curves}
\end{figure}

In the horizontal case, HTCs in the region 
\SIrange{900}{1100}{\watt\per\meter\squared\per\kelvin} were needed to reach the
measured outlet temperatures of approximately $\SI{352}{\kelvin}$. 
In the vertical case, HTCs in the region 
\SIrange{500}{900}{\watt\per\meter\squared\per\kelvin} were needed to reach the
measured outlet temperatures of $\SI{338}{\kelvin}$
and $\SI{348}{\kelvin}$.
In both cases,
the Chen correlation predicted HTCs in the region 
\SIrange{1100}{2600}{\watt\per\meter\squared\per\kelvin}, and reached
temperatures of approximately $\SI{364}{\kelvin}$, which is inside the 
liquid--vapor region.
The Gungor--Winterton correlation predicted HTCs in the region
\SIrange{500}{1300}{\watt\per\meter\squared\per\kelvin}, and reached
temperatures of approximately $\SI{348}{\kelvin}$.

\begin{figure*}
  \centering
  \includegraphics[width=0.81\textwidth]{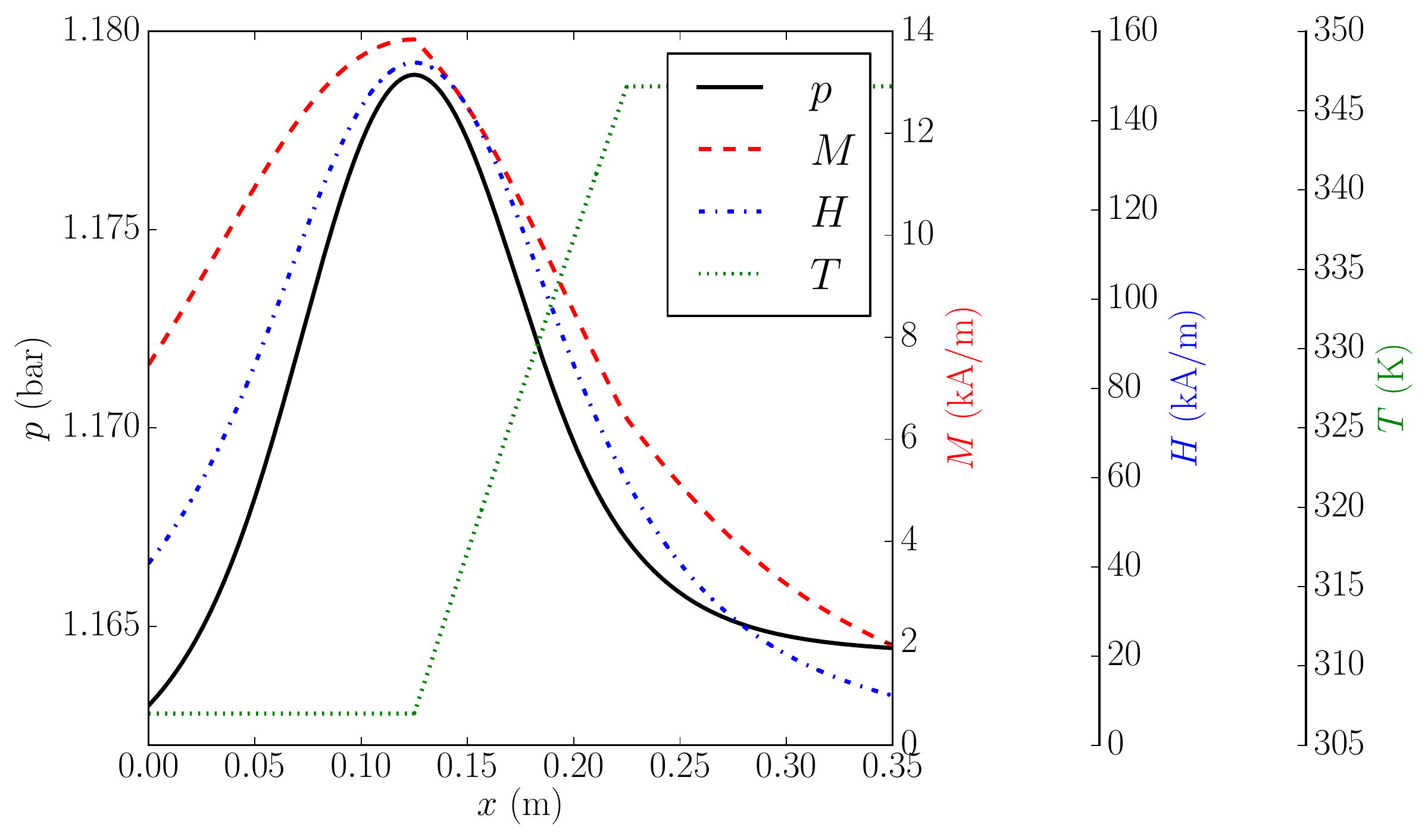}
  \caption{Pressure, fluid magnetization, magnetic field and
    temperature as functions of distance $x$, for a case with $\Delta
    T=\SI{70}{\kelvin}$ and $H_\mathrm{max} =
    \SI{153}{\kilo\ampere\per\meter}$.}
  \label{fig:plot_x}
\end{figure*}

\cref{fig:mag_curves} and~\ref{fig:plot_x} 
illustrate how the magnetization of the fluid
changes through the rig, and how this leads to a net force.
\cref{fig:mag_curves} shows 
the particle magnetization curves at 
the inlet and outlet temperatures, together with the 
$(M,H)$ path taken along the pipe.
\cref{fig:plot_x} shows the total magnetization as a function
of pipe position $x$, together with $p$, $H$ and $T$.
To the left of the solenoid center ($x < \SI{0.125}{\meter}$), the field
gradient $\pd{H}{x}$ and magnetization $M$ are positive, which leads
to a positive magnetic force $f^\text{mag} = \mu_0 M
\pd{H}{x}$. Through the heating element ($\SI{0.125}{\meter} < x <
\SI{0.225}{\meter}$) the temperature increases, which reduces the
fluid magnetization. On the right side of the solenoid center ($x >
\SI{0.125}{\meter}$) the field gradient is negative, but due to the
reduced magnetization, the negative force is less than the positive
force on the left side. This then leads to a net positive pressure
difference $\Delta p$. 

\section{Discussion}
\label{sec:discussion}

Without a symmetry-breaking temperature gradient, the solenoid only provides
equal forces on each of its sides, pointing inwards. This would only lead to 
an internal pressure increase (and slight compression), and no net pressure
increase from one side to the other.
When combining \cref{eq:force_term_mag,eq:mag_f_from_mag_p},
one sees that the local magnetic force is essentially the 
product of $M_\partic$, $\alpha_\partic$, and $\dd H/\dd x$. 
To achieve net pumping action, the magnitude of this force must be
decreased on one side of the solenoid compared to the other.
There are two main mechanisms for achieving this: 

First, the temperature sensitive particle magnetization leads to a decrease of
$M_\partic$ as the temperature increases.
Second, a temperature increase leads to a decrease in the total density 
due to thermal expansion of the base fluid. Due
to conservation of mass in steady state, i.e.\ constant $\rho v$,
the velocity then increases during heating. According
to~\cref{eq:flow_steady_pmass}, this must lead to a decrease in 
$\alpha_\partic$ ($\rho_\partic$ is constant), as the particle 
distribution is stretched thinner by the velocity increase. This effect
can be very pronounced if there is a change from a liquid to a
liquid--vapor state across the solenoid.

Both the above mechanisms are dependent on temperature, and thus the amount 
of heat transferred into the fluid from the pipe walls. 
As in any one-dimensional flow model, for a given temperature difference,
this depends on the HTC 
(see~\cref{eq:source_term_heat}).

This dependence was made clear in the attempt to validate against 
experimental data.
From \cref{fig:deltap_vs_Hmax_0deg_70K} and~\ref{fig:deltap_vs_Hmax_90deg_70K},
it appears that the model is able to successfully predict the 
performance of the thermomagnetic pump, but only accurately if the
prediction of temperature increase is good. As seen, in the simulations
with a constant HTC adjusted to reach the correct outlet temperature,
the predictions for $\Delta p$ are quite close to the measured ones. 
They are not in perfect agreement, but that may come from the fact that 
the whole temperature profile across the solenoid is important, not just
the outlet temperature. The temperature at the right end of the solenoid
is especially crucial, and as seen in \cref{fig:iwamoto_rig}, this is
only halfway across the heater. The actual temperature profile is not
likely to be equal to the one stemming from a constant HTC, even though they
share the same outlet temperature.

In a usage scenario for e.g.\ application design and optimization,
the temperature profile across the solenoid is most likely unknown, and
must be predicted through using a correlation for the heat transfer
coefficient. 
As we see from 
\cref{fig:deltap_vs_Hmax_0deg_70K} and~\ref{fig:deltap_vs_Hmax_90deg_70K}, the
two correlations tested here both give $\Delta p$ in the correct order 
of magnitude, but differ considerably from both each other and the data. 
The results also show the very large sensitivity of $\Delta p$ to uncertainty
in the HTC. 

As we see from the gray bands, a $\pm 25\%$ uncertainty in the HTC 
will lead to an uncertainty of approximately $\pm 50\%$ in $\Delta p$. 
Combined with the fact that such correlations may have much higher
uncertainties in some areas, especially when extrapolating from the
data it is based on, it is apparent that careful use of appropriate
HTC correlations is crucial when assessing the performance of a 
thermomagnetic pump.

In the validation cases used here, the heat transfer was 
of the boiling kind. It was assumed that this effect would 
dominate over any effects of nanoparticles, and thus conventional correlations
for boiling heat transfer could be used. Since the predictions from using
two different such correlations bracket the experimental data, it appears
that this assumption was reasonable. 
However, for laminar flow without boiling, significant Nusselt number 
enhancements from the presence of nanoparticles are 
present~\cite{Kakacc09, Xuan00_2, Maiga05}, 
and correlations for these effects are not nearly as
mature. High uncertainties must then be expected.

In vertical cases, it is worth noting that $\Delta p$ is the result of 
two competing effects: The positive contribution is the same as the one measured
in horizontal cases. The new negative contribution from the magnetic field
is due to the slight densification of the fluid caused by 
the pressure increase inside the solenoid (see \cref{fig:plot_x}).
The magnetic field thus increases the hydrostatic pressure over the fluid
column, decreasing $\Delta p$. Attempting to assist a natural convection
loop with thermomagnetic pumping may thus have a detrimental effect
if the solenoid is placed on a vertical section.

It is also worth noting that the design in \cref{fig:iwamoto_rig} is not
optimal for obtaining maximal $\Delta p$ from the given magnetic field. 
To achieve the largest symmetry breaking of the magnetic force, the
temperature difference between the ends of the solenoid should be as
large as possible.
In terms of \cref{fig:mag_curves}, the net pumping action achieved is
loosely given by the area inside the loop. 
The vertical distance between the red squares is particularly important, as
it shows the difference between the $M_\partic$-factor in the peak
rightward force and the peak leftward force. 
If the temperature were allowed to reach the outlet temperate before 
reaching the right end of the solenoid, the loop would be wider, and 
one would get more pumping action from the same solenoid and heat source.

The benefits of using a model of the kind presented here 
lie in its expected validity across a large range of parameters, 
which is very useful for process design and optimization. 
First of all, the flow equations 
\eqref{eq:flow_steady_pmass}--\eqref{eq:flow_steady_h} are 
generally applicable to any one-dimensional pipe configuration,
no matter if the base fluid is single-phase or two-phase, as long as 
the no-slip assumption holds.

The crucial model for the ferroliquid magnetization, although complex,
is physics-based and contains parameters which all carry physical meaning. 
The model retains the two critical features, S-shaped saturable 
response to external fields and reduction towards the Curie temperature, 
for any set of valid parameters. In contrast, a linearized model would
have to fit new parameters each time the field or temperature changes 
considerably. Combined with the general equation for the solenoid field, 
this ensures a wide range of validity for the magnetic force term.

The thermodynamic model has the big advantage of being based on
an equation of state for the base fluid, 
and thus takes advantage of the very large body of research on 
thermodynamic equations of state, which is not specific for nano/ferrofluids.
This offers a big advantage over linearized or constant-property models, 
as one may perform design and optimization over a variety of 
pressure and temperature ranges with a number of base fluid components,
without having to look up new model parameters.
An equation of state enables the prediction of a large number of thermodynamic
properties, given temperature and pressure, 
in an internally consistent way~\cite{Michelsen07}.
These properties, needed in the flow equations, as well as in various 
HTC correlations, 
include density (with compressibility and thermal expansion), 
heat capacity, enthalpy, latent heat of vaporization, critical pressure, 
bubble line, and liquid--vapor equilibrium. 
Implementations often include parameters 
for many possible components, and
their interactions, so that one may
change the base fluid components on-demand during usage.
Equations of state are semi-empirical to various degrees, range from 
fast and simple to accurate and slow, and any of them may be used with
this model.

A disadvantage is that equations of state are mainly fitted to
equilibrium data, and thus can only reliably represent equilibrium states.
The flow model can thus not account for the presence of vapor in 
subcooled boiling, besides by using an enhanced heat transfer coefficient,
as is done in this work.

Overall, large flexibility and range of validity is expected from a model
of the kind presented in this work. However, it is clear that there is a
major sensitivity to uncertainties in the heat transfer coefficient, as 
would be the case for any one-dimensional heat transfer model with unknown 
temperature profiles. 
Other uncertainties and simplifying assumptions are obviously 
also present, but likely overshadowed by the above.

\section{Conclusions and further work}
\label{sec:conclusions_and_further_work}

We have presented a one-dimensional multi-phase flow model for simulating thermomagnetically pumped ferrofluid flow and heat transfer. The model includes a thermodynamic model which is a combination of a simplified particle model and thermodynamic equations of state for the base fluid. This enables a consistent treatment of partial vaporization of the base fluid, as could result from heat transfer to the ferrofluid. The model also includes a ferrofluid magnetization model, capable of taking into account effects of non-uniform particle size distributions. This magnetization model was shown to accurately reproduce experimental data on ferrofluid magnetization and its dependence on temperature and applied magnetic field. It was also shown that the presented multi-phase flow model was capable of predicting the thermomagnetic pumping mechanism, by which the temperature-dependence of the ferrofluid magnetization led to a net pumping pressure difference, and that the predicted pumping performance was in agreement with the experimental results from~\citet{Iwamoto11}.

It was, however, revealed that the predictions of thermomagnetic pumping are highly sensitive to the temperature profile in the pipe. Using good
correlations for the heat transfer coefficient is therefore absolutely critical, and care must be taken to assure that a correlation is applicable
to a given case. Due to this sensitivity, further research on the effects of nanoparticles on the heat transfer coefficient is called for.

Due to the generality of the models for thermodynamics and magnetization, the proposed model should have a large range of validity, given accurate heat transfer predictions. It could therefore be used for model-based optimization of designs for ferrofluid cooling devices over a wide range of parameters in further research.

Further validation of the model for additional cases would be beneficial,
such as non-boiling heat transfer, and saturated liquid--vapor flow further
inside the solenoid. However, this requires additional
well-documented experiments.

\section*{Acknowledgements}
The research project is funded by the Blue Sky instrument of SINTEF
Energy research through a Strategic Institute Programme (SIP) by the
national Basic Funding scheme of Norway. We would like to thank
Assistant Professor Y.~Iwamoto at Doshisha University, Japan, 
for fruitful discussions, and for providing valuable
information about the experiment in \cite{Iwamoto11}.


\section*{References}
\bibliographystyle{elsarticle-num-names} 
\bibliography{literature}

\end{document}